\documentclass[twocolumn]{aastex7}

\usepackage{fontawesome}
\usepackage{CJK}
\usepackage{lipsum}
\usepackage{xcolor}

\usepackage{wasysym}

\usepackage{graphicx}
\usepackage{subcaption}

\usepackage{amsmath}

\usepackage{comment}
\begin{document}
\title{ From Underground Oceans to Continents: A Glimpse into the Water Inventory on Rocky Planets using Host Star Abundances}


\author[0000-0001-8153-639X]{Kiersten M. Boley}
\altaffiliation{NASA Sagan Fellow}
\affiliation{Earth \& Planets Laboratory, Carnegie Institution for Science, Washington, DC 20015, USA}
\email{kboley@carnegiescience.edu}

\author[0000-0001-5753-2532]{Wendy R. Panero}
\affiliation{Division of Earth Sciences, U.S. National Science Foundation, Alexandria, VA, 22314, USA}
\email{}

\author[0000-0001-8926-4122]{Francesca Miozzi}
\affiliation{Earth \& Planets Laboratory, Carnegie Institution for Science, Washington, DC 20015, USA}
\email{}

\author[0000-0002-2240-0334]{Ashika Capirala}
\affiliation{Department of Earth, Atmospheric, and Planetary Sciences, Purdue University, West Lafayette, IN, 47907, USA}
\email{}

\author[0009-0008-2801-5040]{Johanna K. Teske}
\affiliation{Earth \& Planets Laboratory, Carnegie Institution for Science, Washington, DC 20015, USA}
\affiliation{The Observatories of the Carnegie Institution for Science, Pasadena, CA 91101, USA}
\email{jteske@carnegiescience.edu}




\begin{abstract}
The amount of surface water is thought to be critical for a planet’s climate stability and thus habitability.  However, the probability that a rocky planet may exhibit surface water at any point its evolution is dependent on multiple factors, such as the initial water mass, geochemical evolution, and interior composition. To date, studies have examined the influence of interior composition on the water inventory of the planet or how surface oceans may be impacted by planet topography individually. Here, we provide the first exploration on the impact of interior composition, topography, and planet radius on the water inventory of rocky planets using a sample of 689 rocky planets with spectroscopically derived stellar abundances from APOGEE and GALAH. We find that the oxidation state of the mantle (FeO content) significantly impacts the mantle water storage capacity and potential for surface flooding. For an FeO $\sim$11 wt\%, the water storage capacity of a 1 M$_\oplus$ is 2 times that of Earth, indicating that the oxidation state may reduce the amount of surface water. We quantify the impact of topography on seafloor pressures, showing that flat topographies are more likely to be flooded for all planet compositions and radii. We also find that Mars-like topographies are more likely to have seafloor pressures that may form high-pressure ice, reducing seafloor weathering. Thus, for the first time, we show that the composition and topography of the mantle influence the water inventory of rocky planets. 

\end{abstract}

\section{Introduction} \label{intro}
 
The water inventory of rocky planets significantly impacts the long-term evolution and habitability from the earliest stages of planet evolution. It influences the atmosphere, oceans, and interior dynamics beginning with the magma ocean phase. As rocky planets transition through a global magma ocean phase \citep{Elkins-Tanton2012a,deVries2016,Schaefer2018, Chao2021}, volatiles species (e.g., H$_2$, H$_2$O, CO$_2$) captured during formation readily dissolve within magma \cite[e.g.,][]{Solomatova2020}. While some rocky planets may become perpetual lava worlds due to their proximity to their host star, the majority will cool. The volatiles dissolved within the magma become super-saturated and outgas as the temperature of the planet decreases \citep{ElkinsTanton2008,Salvador2017,Miyazaki2022,Salvador2023}. Once the surface has cooled, liquid water may begin to form on the planet's surface. However, long-term surface water is not guaranteed, as it is a function of the interior and surface conditions of the planet  \cite[e.g.,][]{Kasting1993,ji2023, Guimond2023}. The feedback mechanisms between the atmosphere and interior depend on temperature and composition \citep{Foley2016,Byne2023}. Cold planets feature slow solid-solid interactions between rock and ice layers \citep{Byne2023}. In contrast, lava worlds likely have efficient mass transfer between the surface and atmosphere, making them more likely to equilibrate \citep{Lichtenberg2021,Lichtenberg2021b}. However, cool planets with liquid surface water, such as Earth, have more complex atmosphere-interior interactions as the water facilitates geochemical interactions between the rock, ocean, and atmosphere \cite[e.g.,][]{Foley2016,Kite2018}.

For a rocky planet to have habitable conditions, surface water and geochemical cycles are thought to be essential for climate stability through the regulation of atmospheric CO$_2$, a major greenhouse gas\cite[e.g,][]{Walker1981, Kasting1993,Lammer2009,Irwin2020}. Negative feedback mechanisms such as carbon-silicate weathering are necessary to perturb the climate and reduce CO$_2$ in the atmosphere, resulting in a stable climate and reduced surface temperature \citep{Walker1981, Kasting1993, Foley2016,Foley2018}. On Earth, plate tectonics facilitate the CO$_2$ cycle by driving processes (i.e.,  uplift, orogeny, etc.) that enhance weathering \citep{Kasting2003, West2012}. Plate tectonics may affect the topography of the planet, leading to increased weathering due to the larger area of exposed rock \citep{Foley2015}. However, tectonics are only one mechanism that may regulate climate and prevent a runaway greenhouse planet. Stagnant-lid planets may also have temperate climates given sufficient water and exposed land \citep{Foley2018}. Ocean planets, rocky planets with complete surface flooding, likely do not exhibit geochemical cycles due to the mass of the ocean \cite[e.g.,][]{Foley2015}. Even so, they may also maintain a temperate climate given a small C/H in the atmosphere and ocean \citep{Kite2018, Hayworth2020}. Therefore, constraining the maximum amount of surface water that a rocky planet may have is critical to understanding whether it may host conditions for a stable climate.  

Planet topography impacts the land fraction available for silicate weathering for a given amount of surface water and is influenced by impacts, geodynamics, and planet mass \cite[e.g.,][]{Tenzer2015}. Compared to Earth-mass planets, super-Earths are more likely to become ocean planets with a sufficient amount of surface water due to the relationship between mass, surface gravity, and density.  Planet surface gravity scales as $R^{-2}$ with increasing mass, whereas the density scales as $R^{-3}$. Because of this scaling relation, surface gravity increases at a faster rate than density with increasing mass. This means that small changes in mass have a greater impact on planet topography. Given that super-Earths are more massive than Earth-like planets, they likely have higher surface gravities and thus flatter topographies. As a result, they may be more likely to produce shallow ocean basins (i.e., flat topographies), leading to a high probability of complete surface flooding. However, the flooding potential is dependent on the fraction of water stored within the mantle \citep{Panero2020}.

The transition zone, a layer that divides the upper and lower mantle, is a major water reservoir due to its mineralogy. This region begins at $\sim$12-15 GPa (where olivine no longer forms) and ends at $\sim$22-23 GPa (where perovskite-structured bridgmanite begins to form). The fundamental changes in the structure of major mineralogical components of the mantle within the transition zone affect the overall water storage capacity. In this region, water can be stored as hydroxyl groups (OH$^-$) in defects within the crystal structure of nominally anhydrous minerals (NAMs). Higher pressure olivine polymorphs, such as wadsleyite and ringwoodite ((Mg,Fe)$_2$SiO$_4$), have high water storage capacities and form in larger volume with increasing Mg/Si \citep{Panero2020,Dong2022, Guimond2023}. Additionally, the FeO content of the mantle is critical to understanding the total water storage capacity of the transition zone. Within the mantle, iron exists as FeO and corresponds to the oxygen fugacity or redox state \cite[e.g.,][]{Ledoux2020, Guimond2023b}. The redox state of the planet (FeO content) affects the pressure at which olivine transitions to wadsleyite and ringwoodite, two major water-storing minerals \cite[e.g.,][]{Katsura1989,Katsura2004}.  The change in pressure and thus the upper boundary (low pressure) of the transition zone impacts the total volume of the transition zone. Temperature also controls the amount of water that can be stored within the minerals; At lower temperatures, the capacity to store water increases due to the increase in defects within the crystal in which (OH$^-$) can be stored\citep{Panero2020}.  Therefore, the transition zone will be able to store more water as the planet cools, which has implications for the long-term evolution of the planet. As properties of the transition zone depend on the chemistry and structure of the mineralogical phases, the planet's composition, mass, and temperature will deeply affect transition zone depth range and water storage capacity.

Host star abundances can serve as a proxy for the composition of rocky planets \citep{Adibekyan2021,Schulze2021, Brinkman2024}. We can use Mg, Si, and Fe stellar abundances to infer the interior compositions and thus the maximum water inventories of rocky planets. However, few studies have investigated the interior conditions that impact water storage and surface water \cite[e.g.,][]{Kodama2019,Guimond2023}. Most previous studies that use stellar abundances to constrain the interior compositions of rocky planets have relied on stellar abundances from a variety of sources or catalogs from various instruments and/or analyses. Given that each instrument has systematic offsets, and several studies have shown that these offsets can be significant \cite[e.g.,][]{Soubiran2022}, a heterogeneous sample is less reliable in determining differences within a given population of planets. Without accounting for systematic offsets within each spectroscopic study, non-physical trends may be introduced within the data. With large spectroscopic surveys, such as APOGEE \citep{Sharma2018} and GALAH \citep{Buder2025}, homogenous samples of stellar abundances are available and can give us insight into how planetary systems vary by population.

In this paper, we construct sample of planets with stellar abundances derived from APOGEE \citep{Sharma2018} and GALAH \citep{Buder2025} and account for offsets between the two catalogs. We then use this sample to (1) determine the maximum water capacity of the transition zone as a function of Mg/Si and FeO content, (2) calculate the surface flooding using solar system topographies, (3) compare the maximum water inventory of known planets. This study builds the framework necessary to determine whether surface flooding will occur for a given initial water mass for specific planet compositions. We quantitatively address the Mg/Si ratio and FeO wt\% that produces the highest water storage capacity within the transition zone. We expand on previous studies that consider water storage capacities \cite[e.g.,][]{Guimond2023} by addressing the impact of planet composition on surface flooding and seafloor pressures using solar system topographies.

The outline of our paper is as follows.  Section \ref{s:Method} provides a description of our modeling approach, which outlines all methodology used in calculating known planet compositions (\S \ref{ss:CalculatingKP}),  water storage capacities (\S \ref{s:methodWSC}) and surface flooding (\S \ref{s:SF}). Section \ref{s:KP} discusses our planet sample selection criteria. In Section \ref{s:results}, we discuss the resulting water storage capacities (\S \ref{s:WSC}) and surface flooding (\S \ref{ss:SF}) of super-Earths by planet composition. Here, we also show the water storage capacities (\S \ref{ss:KPWSC}) and impact of surface flooding on seafloor pressure (\S \ref{ss:KPSF}) of our planet sample. We discuss the limitations of our model in Section \ref{s:Discuss} along with additions that will be impactful in future studies before concluding in Section \ref{s:summary and conclusions}. 


\section{Method} \label{s:Method}

The base of our planet interior model uses \texttt{ExoPlex}, a thermodynamically self-consistent planet interior software \citep{2018NatAs...2..297U, Unterborn2019, Unterborn2023}. \texttt{ExoPlex} solves five coupled differential equations to model the mineralogy and thermodynamics of rocky planets. We construct a water storage capacity module to augment this software using methodology to calculate water storage capacities from ab initio calculations in \cite{Panero2020} and \cite{Panero2004}. This module specifically considers the upper limit of water that may be stored within a rocky planet mantle, assuming a spherically symmetric planet for simplicity. 

\texttt{ExoPlex} relies on a fine mesh grid approach to calculate the stable mantle mineral assemblage for a given pressure and temperature from \texttt{Perple\_X}~\citep{Connolly2009} grids. To determine the impact of FeO within the mantle on water storage capacity (\S \ref{s:WSC}), we fix the molar ratios to Earth-like values: Ca/Mg = 0.05, Al/Mg = 0.09, Fe/Mg = 0.9 \cite[e.g.,][]{McDonough2003, Unterborn2023}. We allow Mg/Si to vary as the water storage capacity is dependent on mineralogy and is particularly sensitive to changes in Mg/Si \citep{Panero2020}. Additionally, we allow the FeO content within the mantle to vary. We assume a range of FeO contents from 1-25 wt\% using the maximum FeO content in bodies within the solar system (i.e, $\sim$18-20 wt\% ) to inform FeO contents \citep{Stolper1977,Surkov1984,Surkov1986,Workman2005,Stanley2011,Nittler2018,Khan2022}.  

\subsection{Calculating the Interior of Known Planets} \label{ss:CalculatingKP}
When evaluating known planets in \texttt{ExoPlex}  (\S \ref{ss:KP}), we input their host star molar ratios of Si/Mg, Fe/Mg, Ca/Mg, Al/Mg, which have been shown to match the composition of rocky planets \cite[e.g.,][]{Schulze2021,Brinkman2024}. Given that Al/Mg and Ca/Mg ratios do not significantly impact the water storage capacities of rocky planets, we assume Earth-like ratios for Al/Mg and Ca/Mg when unavailable of 0.09 and 0.07, respectively.  If the planet has a measured mass, we preferentially use mass in conjunction with its radius and host star abundances to calculate the planet's composition in \texttt{ExoPlex}. Otherwise, the planet radius and stellar abundances are used to calculate the mass that corresponds to the measured radius. Using the molar ratios and radius (or mass) , ExoPlex determines the core mass fraction and mineral composition of the mantle from which the water storage capacities may be calculated from \citep{Unterborn2019,Unterborn2023}.

\subsection{Calculating Water Storage Capacities} \label{s:methodWSC}
We consider the effect of composition on the water storage capacity of transition zone minerals, as it dominates the water storage capacity of the mantle of rocky planets \citep[e.g.,][]{Guimond2023}. Within the mantle, nominally anhydrous minerals store water via hydroxyl groups (OH$^-$) instead of H$_2$O. Minerals form a crystalline structure; however, there are defects or vacancies within this structure where OH$^-$ may be stored. There are a variety of defect mechanisms that may lead to water storage. Within this study, we follow the methodology of \cite{Panero2020}, considering the following defect mechanisms: $V_{Mg}^{''}+2H^\bullet$, $V_{Si}^{''''}+4H^\bullet$, $Al_{Si}^{'}+H^\bullet$, $Fe_{Si}^{'}+H^\bullet$. These mechanisms are described using the Kr{\"o}ger-Vink notation, in the form of $V_M^C$ where V  denotes crystallographic vacancies, M for metal species (Mg, Fe, or Ca) occupying lattice site, “$\bullet$” and “ ' ”  to denote a net positive and negative charge, respectively. For example, within (Mg,Fe)$_2$SiO$_4$, the defect mechanism $V_{Mg}^{''}+2H^\bullet$ corresponds to the loss of one Mg, leaving a vacancy denoted by $V_{Mg}^{''}$ and reducing the overall charge by negative two. To balance the charges, two H replace the vacancy caused by the loss of one Mg, resulting in the mineral being neutrally charged. Instead of the defect mechanism being a vacancy, it can be caused by the inclusion of an element. Using the previous example of (Mg,Fe)$_2$SiO$_4$, the defect mechanism $Al_{Si}^{'}+H^\bullet$ would have one Al replace the vacancy left by removing Si.

To determine the internal energy of each phase and defect structure as a function of volume, we calculate the energetics of the $V_{Mg}^{''}+2H^\bullet$, $V_{Si}^{''''}+4H^\bullet$, $Al_{Si}^{'}+H^\bullet$, $Fe_{Si}^{'}+H^\bullet$ defects for the following minerals that are present within the transition zone: wadsleyite, ringwoodite,  akimotoite, davemaoite, periclase, stishovite, and garnet (see Table \ref{tab:Minerals}).  Above a mineral's saturation limit, buoyant, free water is released. As our objective is to assess the storage capacity of the solid planetary interior, we address the water storage capacity of nominally anhydrous minerals below or at their saturation limit. The distribution of water between solid phases can be addressed through defect-specific reactions in the solid state. For example, the reaction of the defect $V_{Si}^{''''}+4H^\bullet$ between two phases with a constant H content written as

\begin{table}[t]
\centering
\begin{tabular}{ccc}
\hline
Mineral & Composition & Location\\
\hline
\hline
Olivine & $\alpha$-(Mg,Fe)$_2$SiO$_4$ & Upper Boundary\\
C2/c pyroxene &(Mg,Fe,Ca)$_2$Si$_2$O$_6$ & Upper Boundary\\
Wadsleyite & $\beta$-(Mg,Fe)$_2$SiO$_4$ & TZ\\
Ringwoodite & $\gamma$-(Mg,Fe)$_2$SiO$_4$ & TZ\\
Garnet &X$_3$Y$_2$(SiO$_4$)$_3$$^*$ & TZ\\
Davemaoite&Ca(Si,Ti)O$_3$ & TZ\\
Akimototite & (Mg,Fe)SiO$_3$& TZ\\
Periclase& (Mg,Fe)O& TZ\\
Stishovite&SiO$_2$ & TZ \\
Bridgmanite&(Mg,Fe)SiO$_3$& Lower Boundary \\
\hline
\end{tabular}
\caption{Mineral compositions that form in the transition zone (TZ) and those that define the upper and lower boundaries. $^*$X may be (Mg,Fe,Mn)$^{+3}$ and Y may be (Al,Fe,Cr)$^{+2}$. \label{tab:Minerals}}
\end{table}

\begin{align}
    Mg_2SiO_4^\alpha+Mg_2Si_{(1-x)}H_{4x}O_4^\beta= M&g_2Si_{(1-x)} H_{4x}O_4^\alpha \notag\\
     &+Mg_2SiO_4^\beta
\end{align}
\\
\\
To calculate the distribution of hydrogen between two phases $\alpha$ and $\beta$, we use Henry's law

\begin{equation}
    K^H_{\alpha/\beta}=exp\left(\frac{\Delta G}{k_BT}\right)
\end{equation}
\\
\\
where $k_B$ is Boltzmann's constant, $T$ is temperature, and $\Delta G$ is the Gibbs free energy of the reaction between phases $\alpha$ and $\beta$ at a given defect with constant H content calculated as 
\begin{equation}
    \Delta G=\Delta H^{RXN}-T \left(\Delta S^\alpha_{config} -\Delta S^\beta_{config} \right)
\end{equation}
\\
\\
where $\Delta H^{RXN}$ is the enthalpy (i.e, the the measurement of energy in a thermodynamic system) of each defect formation. $\Delta S^\alpha_{config}$ is the configurational entropy 
\begin{equation}
    S=k_B ln\left(\Omega_N\right)
\end{equation}
\\
\\
$\Omega_N$ is the number of possible configurations.

We calculate the water storage capacity of the minerals in the transition zone using the calculated distribution of hydrogen between wadsleyite and additional transition zone minerals. We define the upper boundary where olivine no longer forms. However, olivine does not form for all compositions. In such cases, we instead define the upper boundary of the transition zone at pressures at which C2/c pyroxene transitions to wadsleyite or ringwoodite when necessary. The lower boundary is set by the formation of bridgmanite.

\subsection{Surface Flooding} \label{s:SF}

\subsubsection{Initial Water Inventory}

There are two predominant methods in which small planets may obtain water: (1) via deposit of volatiles due to the accretion small bodies, (2) interaction with the primordial H$_2$ atmosphere during formation. The origin of Earth's water is conventionally attributed to the late accretion of small bodies, such as asteroids or comets due to the deuterium/hydrogen (D/H) of the Sun (D/H $\sim 2.0 \times 10^{-5}$) \citep{Gloeckler1998} and small bodies (D/H $\sim2-6 \times 10^{-4}$) \citep{Bockel2012,Sarafian2017} compared to that of Earth's oceans (D/H $\sim1.5 \times 10^{-4}$) \citep{Geiss2003}. Specifically, D/H within Earth's ocean is particularly low when compared all other sources of D/H in the Solar System, including comets \cite[e.g.,][]{Bockel1998,Altwegg2015,Alexander2018}, chondrite meteorites \cite[e.g.,][]{Bockel1998,Altwegg2015,Alexander2018} and water in icy moons \citep{Lis2019}. However, D/H is degenerate with many processes as reactions between H$_2$ and H$_2$O result in D/H of H$_2$O being greater than that of H$_2$. Several studies suggest that present day D/H of Earth's oceans may be also be possible if Earth's water was obtained via an interaction the H$_2$ primordial atmosphere \cite[e.g.,][]{Genda2008,Saito2020, Young2023}. Therefore, we choose to focus on the latter as it is challenging to constrain the amount of water that may be deposited via meteorites due to the complex dynamics of protoplanetary disks. 

Given that we aim to understand the upper limit of water storage and surface water storage, we calculate the maximum H$_2$ atmosphere that is possible for a super-Earth following equation 24 from \cite{Ginzburg2016}, 

\begin{equation}
    f=0.01 \left( \frac{M_p}{M_\oplus}\right)^{0.44}\left( \frac{T_{eq,neb}}{1000 K}\right)^{0.25} \left( \frac{t_{disk}}{1Myr}\right)^{0.5}
\end{equation}

where $f$ is the gas fraction, $M_p$ is the planet mass, $T_{eq,neb}$ is the equilibrium temperature of the surrounding nebula, and  $t_{disk}$ is the lifetime of the disk. We assume the equilibrium temperature to be 1000 K in line with previous studies correlating to in situ formation within the inner disk \cite[e.g.,][]{Ginzburg2016, Young2023}. Given that the median disk lifetime ranges from 1-10 Myr \cite[e.g.,][]{Ribas2015,Richert2018,Briceno2019,Michel2021,Pfalzner2022}, we assume the upper limit of $t_{disk} = 10 Myr$. 

H$_2$ has been suggested to interact with the FeO component of the melt through the reaction FeO + H$_2$ $\Rightarrow$ H$_2$O + FeH during the magma ocean phase. Here, we use this equation to calculate the amount of water produced given the planet mass and FeO in the mantle. We use the solubility limits from Miozzi et al 2025 (submitted) to limit the maximum amount of water that may be dissolved into the melt during the magma ocean phase. Given that the conversion of FeO to H$_2$O is unlikely to be 100\% efficient, we impose an efficiency factor to scale the FeO conversion to that of Earth, reproducing results from \cite{Young2023}.

We set the water inventory within the transition zone to be the maximum water storage capacity directly following solidification. Each planetary Mg/Si ratio and FeO variation corresponds to a unique composition, which will solidify at different temperatures based on its unique melting curve. Given the limited data for melting curves for these compositions, we estimate the solidification temperature using the hydrous solidus curve described in \cite{Boley2023}. Finally, we assume that any water that is not stored in the solid mantle percolates to the surface to become surface water. This gives us the most optimistic scenario to determine the range of planets that may exhibit surface oceans during their evolution.

\subsubsection{Hypsometric Curve}\label{ss:mHypso}
\begin{figure}[!t] 
\begin{center}
\includegraphics[width=\linewidth]{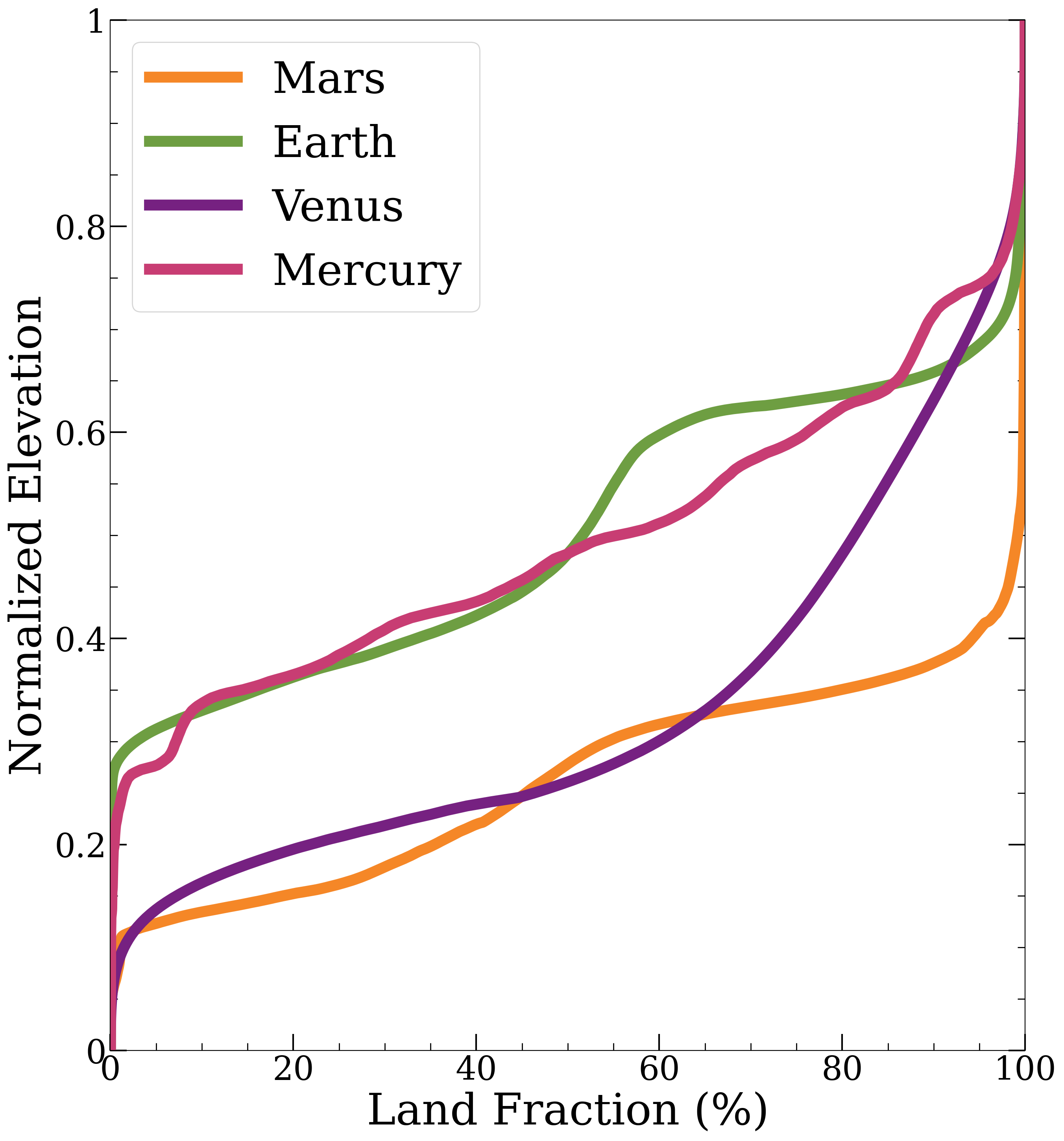}
\caption{The normalized hypsometry of 1 M $_\oplus$ planet calculated using topographic data from Earth (teal)\citep{amante2009}, Mars  (purple) \citep{fergason2018(MARSData)}, Venus (pink) \citep{VenusData}, and Mercury (gray) \citep{MercuryData}. Mercury and Earth have the most shallow topographies. Given the steep topographies of Mars and Venus, they are most useful when considering the surfaces of Earth-like or smaller ($<$ 1 R$_\oplus$) }
\label{fig:Hyps}
\end{center}
\end{figure} 

Within the solar system, rocky planets have varying topographies that are driven by a variety processes. On Earth, the topography is largely a result of plate tectonics \citep{Braun2010,Forte2022}. Similarly, the topography of Mars may have experienced early plate tectonics or regional tectonism and was also affected by volcanic accumulations and impact of meteorites \cite[e.g.,][]{Nimmo2000,Tenzer2015, McCullough2024}. Venus' topography is thought to be caused by internal overturns that result in episodic global resurfacing, given its stagnant lid \citep{Tian2023}, whereas Mercury's topography is thought to be a result of impacts  \citep{Rivera2014}. These processes impact a planet's topography, which influences the depth of its ocean basin. Therefore, it is important to understand how topographies sculpted by different processes may impact the maximum surface water to produce ocean planet and land fractions on partially flooded planets.

To determine the maximum amount of water required to cause surface flooding, we assume topographies of solar system planets to include Earth, Venus, Mars, and Mercury. We quantify the planet topography using a hypsometric curve, where hypsometry describes area distribution at different elevations \citep{Strahler1952,Schumm1956}. 
To calculate the hypsometric curve, we use observational topographical data. We employ the ETOPO5 Global Relief Model \citep{amante2009} for Earth, which includes both topography and bathymetry (i.e., seabed topography). For Mars, we use Mars Orbiter Laser Altimeter (MOLA) data \citep{fergason2018(MARSData)},  Venus Magellan Compressed Mosaicked Image Data \citep{VenusData} for Venus, and the Mercury MESSENGER MDIS Basemap LOI Global Mosaic 166m for Mercury \citep{MercuryData}. For Earth, Mars, Venus, and Mercury, the elevation uncertainty is $\pm$10 m, $\pm$3 m, $\pm$50 m, $\pm$21 m, respectively. For the purposes of this study, the uncertainties are negligible as we aim to understand our Solar System-like topographies may influence surface flooding.
The hypsometric curves are then used to calculate the fraction of land uncovered by surface water ($f_{land}$). For a given mass of surface water, we iterate over the hyposmetric curve to determine the mass of the water required for complete surface flooding  ($M_{flood}$) using, 

\begin{equation}
     M_{flood} =\sum^{n}_{i=1} M_{flood,i}
\end{equation}

where $M_{flood,i}$ is the incremental increase of the flood mass that corresponds to the flood height increment set by the elevation uncertainty multiplied by the water coverage (1- $f_{land}$), the surface area, and water density, 

\begin{equation}
     M_{flood,i} = H (1 - f_{land,i}) \rho_{H_2O} \hspace{1mm}4  \pi (r + H_i - D_i)^2
\end{equation}

where $\rho_{H_2O}$ is the density of water\footnote[1]{We assume the density of pure water ($\rho_{H_2O} \sim 1 \frac{g}{cm^3}$), not accounting for the compression of water at higher pressures.}, $r$ is the planet radius, and $H$ is the flood height increment. $D$ is the land depression caused by the surface water (i.e., isostasy). The topographical map of Earth accounts for isostasy  as Earth's surface oceans depress the land below sea level. To account for isostasy only once, we calculate land depression due to water above sea level for Earth. However, for Mercury, Venus, and Mars, we account for isostasy as they do not have surface oceans causing land depression. We use the following equation for isostasy,

\begin{equation}
    D = \frac{\rho_{H_2O}\times H}{\rho_{mantle}} 
\end{equation}

where $\rho_{mantle}$ is the density of the mantle. We show the normalized hypsometric curves for a 1 M$_\oplus$ planet in Figure \ref{fig:Hyps}.




\section{Planet Sample} \label{s:KP}

\subsection{Acquiring Stellar Abundances}
To determine planet composition, we first construct a stellar sample that accounts for systematic offsets between surveys. We use APOGEE DR17 \citep{Abdurro2022APOGEE} and GALAH DR4 \citep{Buder2025} and apply the following quality cuts as recommended by both surveys:

\begin{enumerate}
    \item \textbf{Signal to Noise:} For GALAH, we use the parameter \texttt{snr\_px\_ccd3} for SNR and require it to be above 30 as suggested in the recommended quality cuts. For APOGEE, we also use the recommended quality cut requiring the \texttt{SNR} flag  to be above 20.   
    \item \textbf{Removing Data with Known Errors:}  We remove all stars with known quality issues in their stellar parameters, and these issues could propagate to inaccurate abundances. 
    \begin{enumerate}
        \item For GALAH, we begin by setting the \texttt{flap\_sp} flag to zero to eliminate entries with identified problems in the determination of their stellar parameters.
        \item For APOGEE, we use the equivalent stellar parameter flag to GALAH and require \texttt{ASPCAPFLAG} to be equal to zero.
    \end{enumerate}
    \item \textbf{Abundance quality flags:} Each catalog contains quality flags for the measured abundances. We exclude stars that have a flag indicating issues with the determination of the [Fe/H], [Mg/Fe], or [Si/Fe] abundances. Given that [Al/Fe] and [Ca/Fe] do not significantly impact the water storage capacity of rocky planets, we do not remove stars with flags on these measurements. However, we remove the flagged [Al/Fe] and [Ca/Fe] data from our calculations. We consider the following flags in each survey:
    \begin{enumerate}
        \item For GALAH, we require the \texttt{flag\_mg\_fe, flag\_si\_fe} and \texttt{flag\_fe\_h} flags to all be equal to zero. While we do not remove stars without [Ca/Fe] or [Al/Fe], we only keep this data for stars with \texttt{flag\_ca\_fe} and  \texttt{flag\_al\_fe} equal to zero. 
        \item For APOGEE, we require the \texttt{MG\_FE\_FLAG, SI\_FE\_FLAG} and \texttt{FE\_H\_FLAG} flags to all be equal to zero. Similar to GALAH, we do not remove stars without [Ca/Fe] or [Al/Fe]. However, we only keep [Ca/Fe] or [Al/Fe] data for stars with \texttt{CA\_FE\_FLAG} and  \texttt{AL\_FE\_FLAG} equal to zero.
    \end{enumerate}
\end{enumerate}
These cuts result in a sample of 1,169,384 stars. We perform a cross-calibration of the surveys following the procedure in \cite{Soubiran2022} to account for systematic offsets in the abundance measurements across the two surveys, where a subsample of stars with abundances in both surveys is used to calculate the offsets between catalogs.

\subsection{Planet Selection}
\begin{figure}[t] 
\begin{center}
\includegraphics[width=\linewidth]{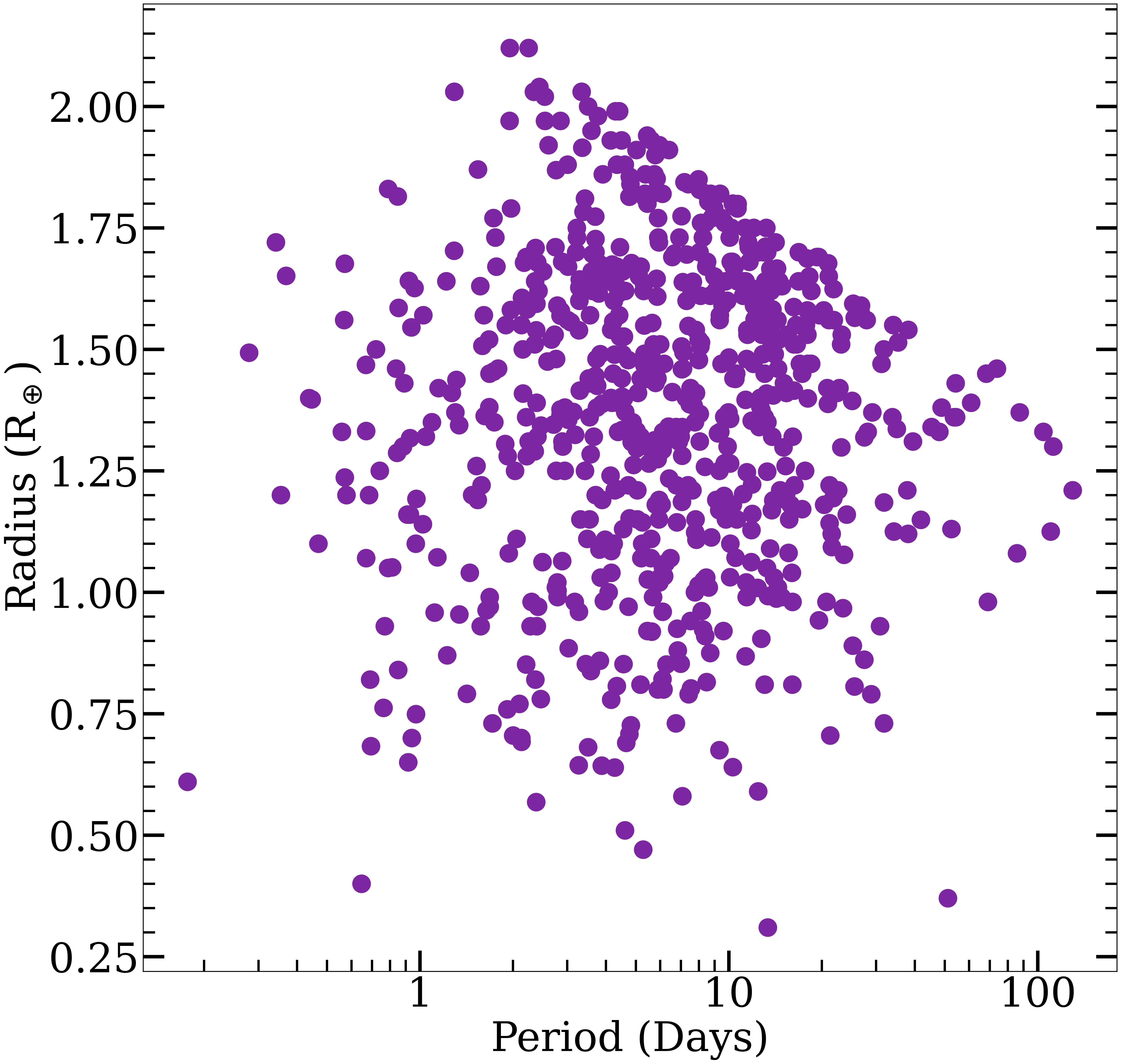}
\caption{Planet sample as a function of radius and period. Our total sample consists of 689 rocky planets. We use a period-dependent radius gap \citep{VanEylen2018, Ho2023} to classify super-Earth and Earth-like planets.}
\label{fig:PlanetSample}
\end{center}
\end{figure}

Once calibrated, we construct our planet sample using the NASA Exoplanet Archive \citep{Christiansen2025}\footnote[2]{accessed on 7/23/25} and applying the following cuts. We initially require all planets to have radius measurements. We remove any planets with flagged or controversial measurements. Given that there are multiple published values for each planet, we keep planets with the least relative uncertainties on radius. To ensure that we only consider rocky planets, we further remove the planets above the radius valley using the period-dependent radius gap that are considered to be sub-Neptunes \citep{VanEylen2018, Ho2023}. Specifically, we use equation 6 from  \citep{Ho2023} to define the radius valley.  Given that the extent to which the radius valley depends on stellar mass is still an active area of study \cite[e.g.][]{Gaidos2024,Ho2024,Parashivamurthy2025}, particularly at low stellar mass ($\lesssim$ 0.7 $M_\odot$), we choose to optimize sample size including potentially rocky planets from equation 6 in \citep{Ho2023}. Once we have our planet sample (Table \ref{tab:planetsamp}), we cross-match the data with our stellar sample.  We then convert the stellar abundances to molar ratios (see Table \ref{tab:sE}) using the procedure described in \cite{Hinkel2014}, where [X/H] is converted to molar ratios using solar abundances from \cite{Lodders2010}. Our final planet sample consists 689 super-Earths, 56 of which have measured masses (Figure \ref{fig:PlanetSample}). Within Table \ref{tab:planetsamp}, we show the range and median of planet and host star properties.

\begin{table}[t]
\centering
\begin{tabular}{cccc}
\hline
Parameter & Range & Median & Units\\
\hline
\hline
Planet Mass & 0.07 -- 9.7 $^*$& 4.44 &$M_\oplus$\\
Planet Radius & 0.31 -- 2.12& 1.61 &$R_\oplus$\\
Stellar Mass & 0.4 -- 1.584 & 0.91 &$M_\odot$\\
Stellar Temperature & 3612 -- 6433 & 5405 & K\\
$[Fe/H] $& -0.82 -- 0.43 & -0.01 & \\
Mg/Si & 0.59 -- 1.71 & 1.01& \\
Fe/Mg& 0.09 -- 1.74 & 0.61& \\
\hline
\end{tabular}
\caption{Planet Sample Parameters: We show the ranges and medians for each parameter range of our sample with stellar properties from APOGEE \citep{Abdurro2022APOGEE} and GALAH \citep{Buder2025} and planet properties from the NASA Exoplanet Archive. $^*$ Note that the range and median derived from the planet masses are limited to 58 planets within the sample with measured masses.  \label{tab:planetsamp}}
\end{table}

\section{Results} \label{s:results}

\begin{figure*}[!t] 
\begin{center}
\includegraphics[width=\linewidth]{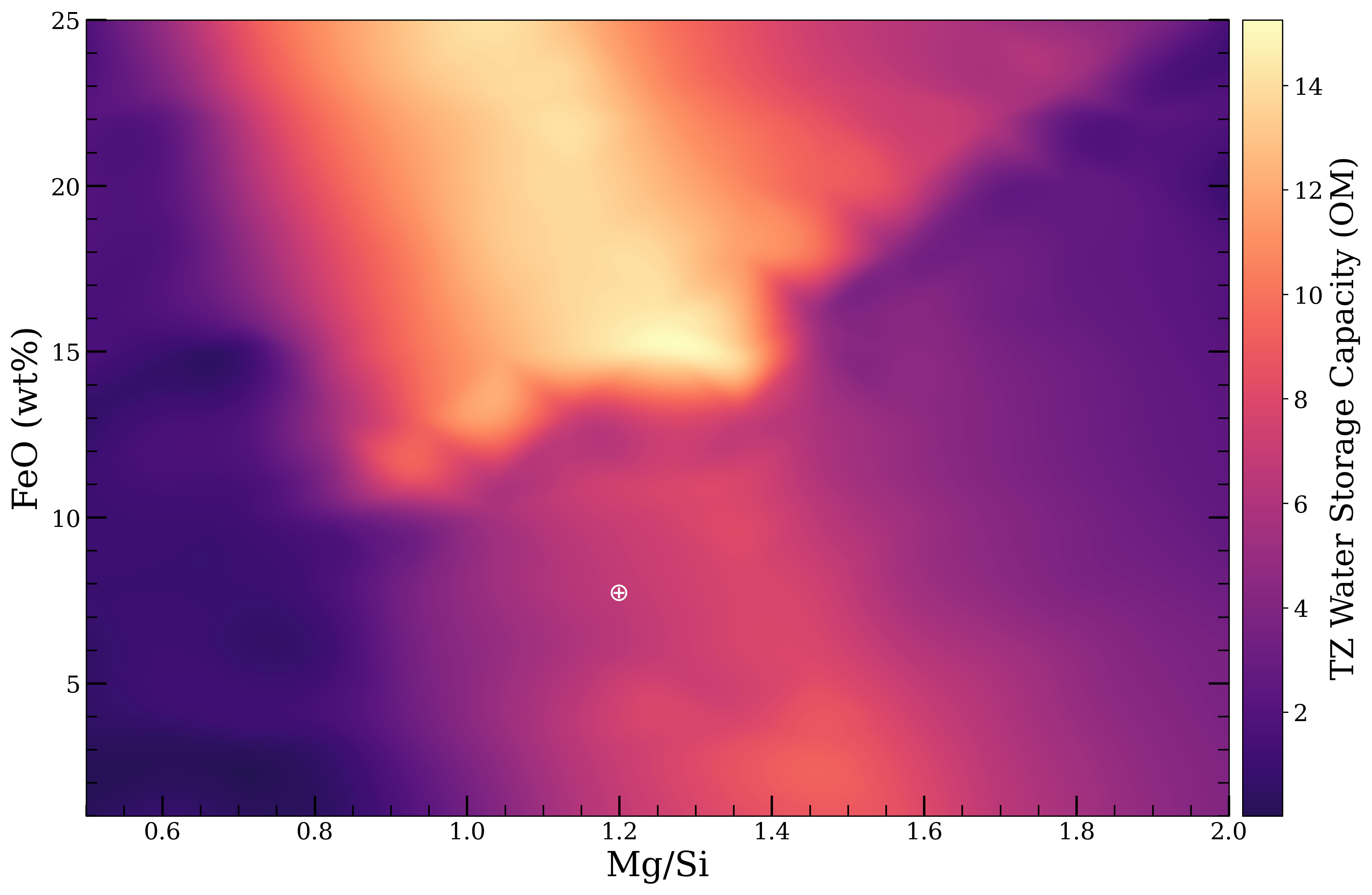}
\caption{Transition zone (TZ) water storage capacity as a function of Mg/Si and FeO for a 1 M$_\oplus$ planet with a mantle potential temperature of 1600 K. We use cubic interpolation between data points for smoothing. The colorbar indicates the water storage capacity in Earth oceans. Earth's composition is indicated by $\oplus$. The TZ water storage capacity increases with FeO. The drastic increase in water storage capacity at FeO $\sim$11 wt\% corresponds with a pressure change in the transition of olivine to wadsleyite.}
\label{fig:PlanetWC}
\end{center}
\end{figure*}

\begin{figure*}[!t] 
\begin{center}
\includegraphics[width=\linewidth]{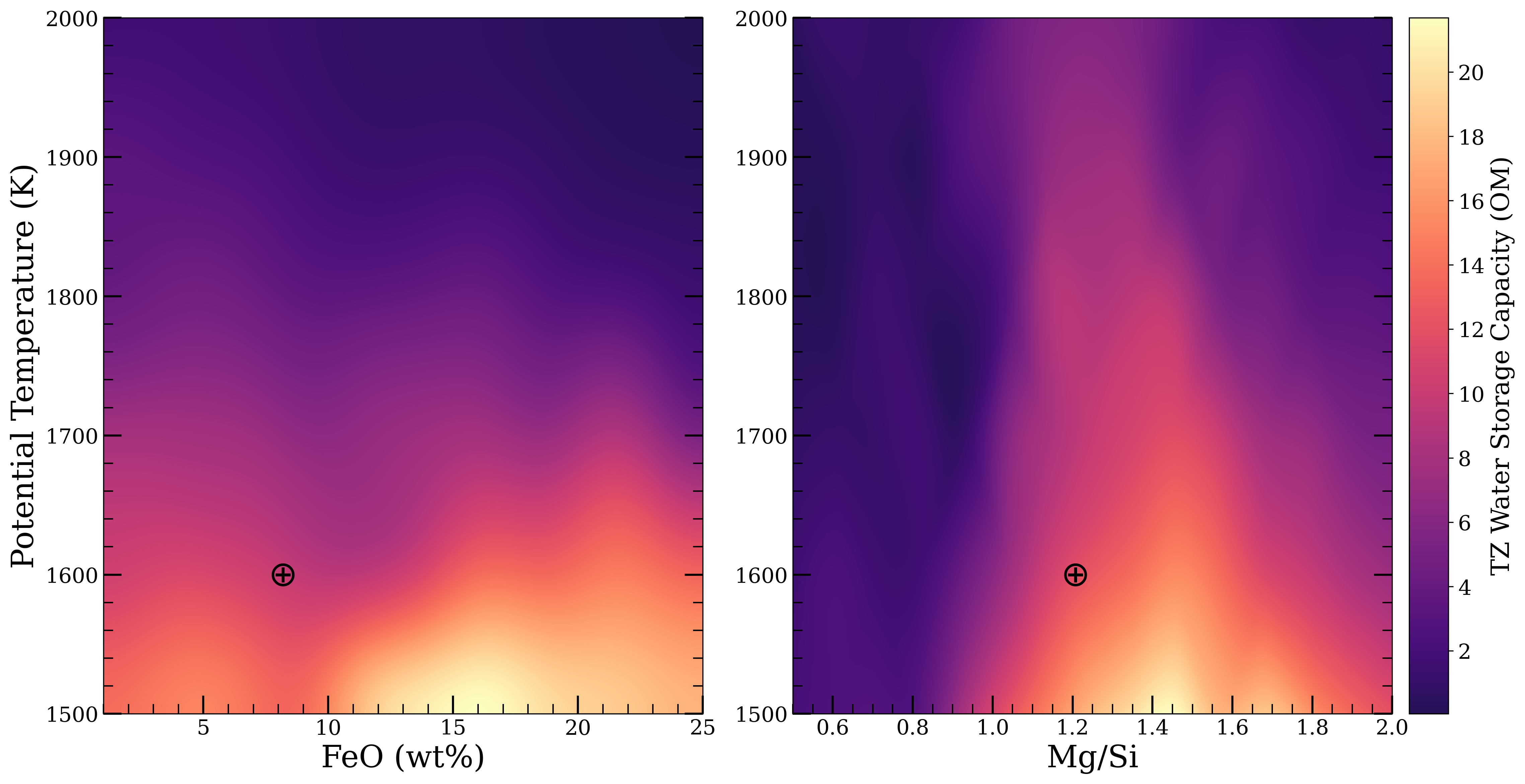}
\caption{TZ water storage capacity as a function of composition and mantle potential temperature assuming a 1 M$_\oplus$ planet. The colorbar indicates the water storage capacity in Earth oceans. We use cubic interpolation between data points for smoothing.\textbf{Right:} We show the variation of water storage capacity with temperature as a function of FeO. Decreasing temperature corresponds with a higher water storage capacity. \textbf{Left:} We show the variation of water storage capacity with temperature as a function of Mg/Si. Generally, (Mg,Fe)$_2$SiO$_4$ minerals (i.e., wadsleyite and ringwoodite) show a greater dependence on temperature than other minerals. }
\label{fig:Temp}
\end{center}
\end{figure*} 
\subsection{Mantle Water Storage Capacity}\label{s:WSC}
The water storage capacity of the transition zone is dependent on the mineralogical assemblage forming for the given Mg/Si ratio and FeO content.Using the methodology described in \S \ref{s:methodWSC}, we calculate the water storage capacity for minerals within the transition zone. Wadsleyite does not form within the transition zone of planets with Mg/Si $\lesssim$0.8 for the majority of compositions considered. However, the minimum Mg/Si for wadsleyite to form decreases as FeO increases. Given that wadsleyite is a major water reservoir within the transition zone, it significantly increases the water storage capacity of the transition zone. Ringwoodite, garnet, and davemaoite form at all Mg/Si compositions with ringwoodite and garnet having the highest water storage capacities, following wadsleyite. Ringwoodite has higher water storage capacities with increasing FeO content that is also linearly dependent with Mg/Si. At FeO $\gtrsim$15 wt\%, ringwoodite becomes the dominate water storing mineral instead of wadsleyite.  Akimotoite, perclase, and stishovite form at various Mg/Si intervals and have little impact on the overall water storage of the transition zone compared to wadsleyite, ringwoodite, and garnet.

Figure \ref{fig:PlanetWC} illustrates the transition zone water storage capacity in terms of Earth Oceans (OM; 1 OM=$1.335 \times 10^{21}$ kg) of a 1 M$_\oplus$ planet with a mantle potential temperature (T$_m$) of 1600 K as a function of Mg/Si and FeO content. Here, we fix the abundances to Earth-like compositions for Ca/Mg, Al/Mg, and Fe/Mg.  Within the parameter space of Figure \ref{fig:PlanetWC}, the water storage capacity peaks for Mg/Si between 1--1.6 for FeO $\gtrsim$11 wt\%. For FeO $\lesssim$11 wt\%, there is a decrease in the water storage capacity with increasing FeO content, which is most prominent in the peak  Mg/Si $\sim$ 1.4.  For example, at Mg/Si= 1.3 and an FeO= 1 wt\% the maximum water storage capacity is $\sim$ 8 OM, whereas at an FeO= 10 wt\% is $\sim$ 7 OM. At FeO $\gtrsim$11 wt\%, there is a significant increase in water storage capacity for Mg/Si between 1--1.6 due to the increase in FeO content, which reduces the pressure at which olivine transitions to wadsleyite and ringwoodite. The lower pressure transition of olivine results in the transition zone having a larger total volume, and thus a higher water storage capacity. At FeO $\gtrsim$11 wt\%, water storage capacity peaks at lower Mg/Si ratios at Mg/Si between 1.2--1.4 compared to the lower FeO region (FeO $\lesssim$11 wt\%) and similarly decreases with increasing FeO content after the initial peak. The increase of water storage is 1.5--2 times the water mass of transition zones with FeO $\lesssim$11 wt\%. 

The water storage is significantly influenced by wadsleyite and ringwoodite. The formation of wadsleyite is dependent on both Mg/Si and FeO. At low FeO content (1-5 wt\%), more wadsleyite forms at lower Mg/Si ratios compared to higher FeO content(6-10 wt\%). For example,  wadsleyite forms at Mg/Si $\lesssim$0.8 at FeO =1 wt\%,. However, this trend "resets" after the FeO$\sim$ 11 wt\% where the overall volume of the transition zone increases due to the lower pressure formation of wadsleyite. At FeO $\sim$10 wt\%, wadsleyite begins to from at Mg/Si $\sim$1.1, but forms again at Mg/Si$\sim$ 0.8 at FeO$\sim$11 wt\%. Ringwoodite also has a linear dependence in its water storage capacity with FeO and Mg/Si. However, it behaves the opposite of wadsleyite. Increasing FeO content causes an increase in the water storage capacity of ringwoodite at lower Mg/Si ratios. For example, the water storage capacity of ringwoodite at Mg/Si=1 and FeO= 5 wt\% is $\sim$ 1 OM, whereas it is $\sim$8 OM at Mg/Si=0.65 and FeO= 25 wt\%.

\subsubsection{Temperature Dependence}\label{ss:Temp}
The effect of temperature on the water storage within the transition zone is important to understand how the water inventory of a planet may vary over its evolution. Previous studies indicate that water storage capacities of individual minerals are sensitive to changes in temperature \cite[e.g.,][]{Panero2020}. As the temperature of the mineral decreases, the vacancies within the lattice increase. The resulting vacancies increase the overall water storage capacity of the mineral. We examine this effect for total water storage within the transition zone of rocky planets.

We vary the mantle potential temperature from 1500 -- 2000 K for the full range of compositions (1 wt\% $\leq$ FeO $\leq$ 25 wt\% and 0.5 $\leq$ MgSi $\leq$ 2) for a 1 M$_\oplus$ planet. We first consider the impact on water storage capacity with temperature as a function of FeO wt\% shown in Figure \ref{fig:Temp} (Left).  As temperature decreases, we find a relatively uniform increase of the water storage capacities across all FeO contents until $\sim$1600 K. Below 1600 K, the water storage capacity increase varies across for FeO above or below $\sim$11 wt\%. At $T_m <$ 1600 K and FeO $>$ 11 wt\%, the water storage capacities increases 1.5--1.7 times that compared to $T_m <$ 1600 K and FeO $<$ 11 wt\%. The significant increase is due to the transition of olivine to wadsleyite. 

In Figure \ref{fig:Temp} (Right), we illustrate the temperature dependence of Mg/Si. Similar to FeO, we find an increase in the water storage capacity as a function of temperature for variations in Mg/Si. The impact of temperature on the water storage capacity is most significant for Mg/Si between 0.9--2, which corresponds to the sensitivity of wadleyite and ringwoodite. At Mg/Si $<$ 0.9, there is still an increase in the water storage capacity; however, it is not as significant given the low abundance of (Mg,Fe)$_2$SiO$_4$ minerals. We find that below Mg/Si $<$ 0.9, the water storage capacities increase by $\sim$2 times the capacity at 1500 K compared to 2000 K. However, the increase is $\sim$5 times the amount at 1500 K than at 2000 K for Mg/Si $\sim$1.4.

\subsubsection{Radius Dependence}\label{s:WSCvT}
\begin{figure}[!t] 
\begin{center}
\includegraphics[width=\linewidth]{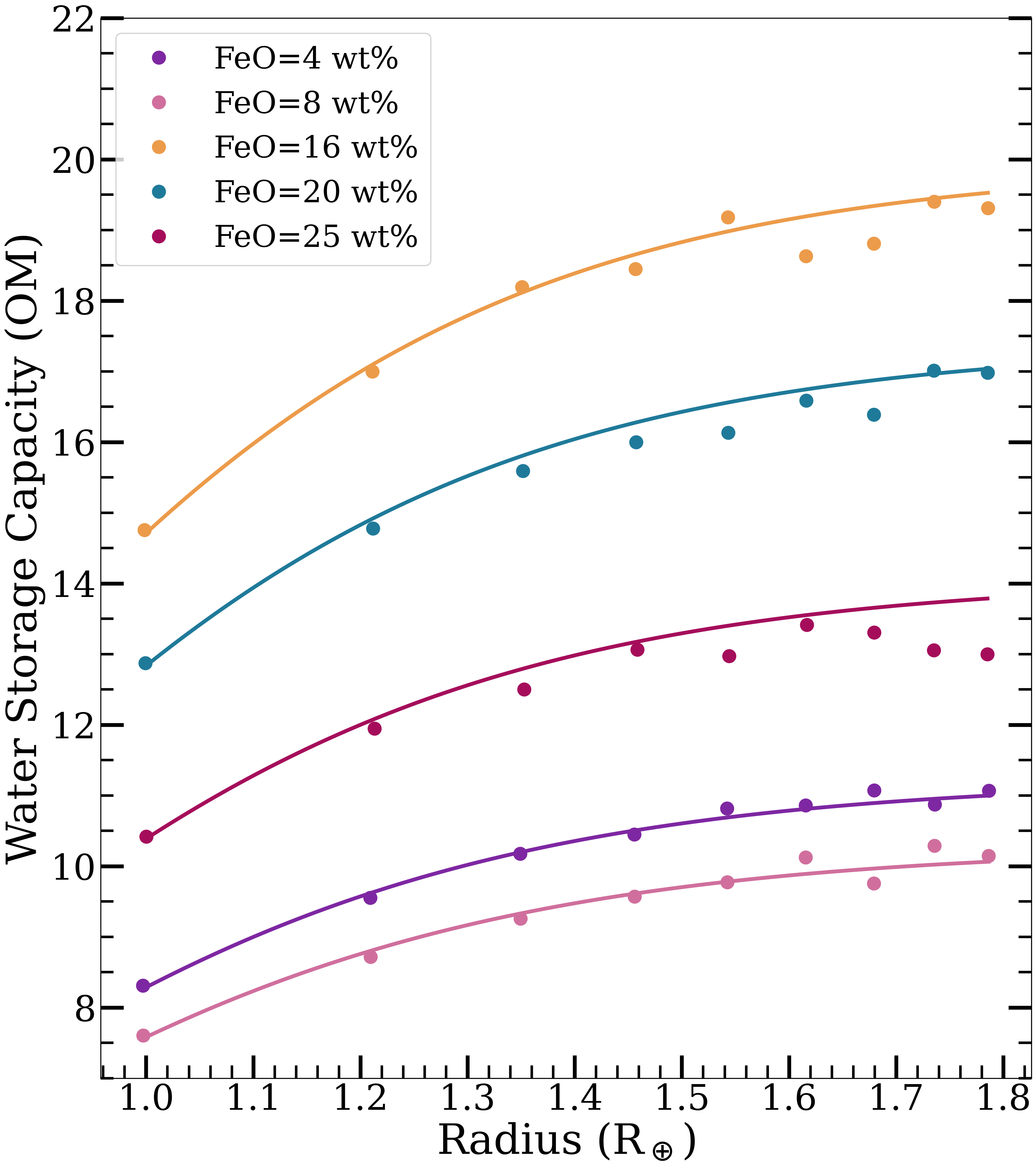}
\caption{Water storage capacity as a function of planet radius and composition, assuming a Mg/Si= 1.2. We show the impact of FeO on the water storage capacities as a function of radius and their corresponding fits given by equation (\ref{eq:WSC}).}
\label{fig:WSCfunction}
\end{center}
\end{figure}

Figure \ref{fig:WSCfunction} shows the relationship between water storage capacity as a function of radius for different FeO content with Mg/Si=1.2 and mantle potential temperature of 1600 K. We find that for fixed Mg/Si and FeO content, the water storage capacity scales similarly as a function of radius. The correlation of water storage with planet radius is largely due to the increase in volume of the transition zone. 

We determine a scaling relation for planet composition and planet radius. We choose a physically-motivated sigmoid to model the water storage capacity as it asymptotes at small and large radii. We do this to ensure at the small  radii limit all values remain non-negative for the water storage capacity. Similarly, at the large radii limit, the model is not divergent yielding infinite values for the water storage capacity.  The water storage capacity sigmoid is of the form: 

\begin{equation} \label{eq:WSC}
    W_{TZ}= w_0  \left( \frac{a}{ 1 + e^{-b(R_p - R_{p,0})}}\right)
\end{equation}
\\
where $W_{TZ}$ is the water storage capacity of the transition zone, $w_0$ is the water storage capacity of the 1 $M_\oplus$ planet of corresponding composition (see Figure \ref{fig:PlanetWC} for values), $R_p$ is planet radius in Earth radii where $R_p \leq 2$ , and $a$, $b$, and $R_{p,0}$  are constants. To determine the best-fit values for $a$, $b$, and $R_{p,0}$ , we employed \texttt{scipy.optimize.curve$\_$fit}. This software relies on a nonlinear least squares method to fit the sigmoid \citep{Vugrin2007}.  We find the best-fit parameters to be $a =1.351 \pm 0.003$, $b = -3.59 \pm 0.08$, and $R_{p,0}= 0.712 ± 0.006$ using 1$\sigma$ uncertainties. 

\begin{figure*}[t] 
\begin{center}
\includegraphics[width=\linewidth]{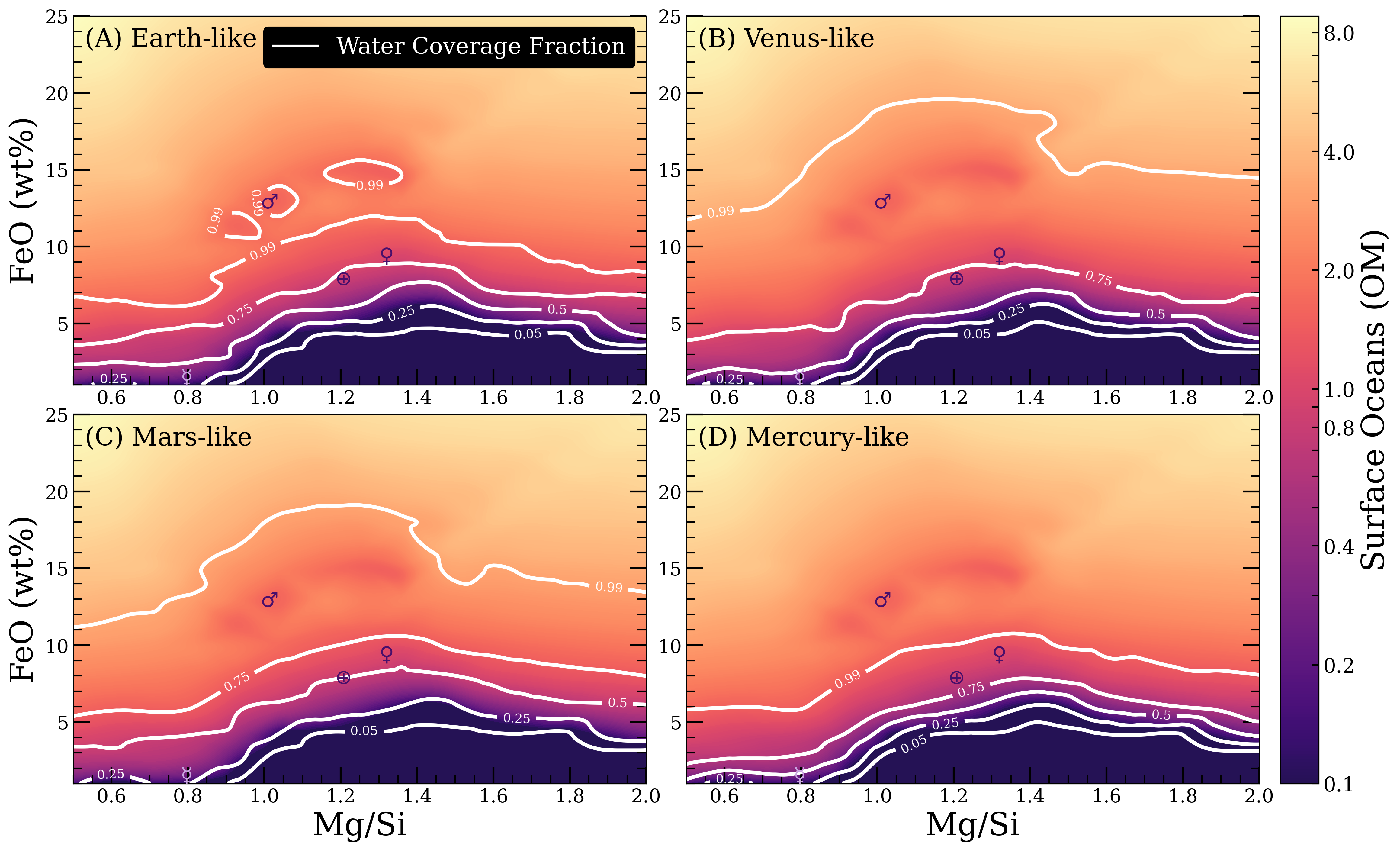}
\caption{We show the total surface flooding for a given topography as a function of Mg/Si and FeO for a 1 M$_\oplus$ planet with Earth-like molar ratios (i.e., Fe/Mg, Ca/Mg, Al/Mg). We use cubic interpolation between data points for smoothing. The colorbar indicates the surface flooding in Earth oceans. We show the compositions of solar system planets with the following notation: Earth ($\oplus$), Venus ($\female$), Mars ($\male$), Mercury ($\mercury$), using compositional values from \cite{Tronnes2019}.  (A)  Earth-like topography shows three distinct "islands" correspond to the increase in water storage capacity. (B) Venus-like and (C) Mars-like topographies show similar structure in the water coverage fraction. (D) Mercury-like topography reaches complete flooding at low OM. }
\label{fig:SurfFlood}
\end{center}
\end{figure*}

\subsection{Surface Flooding} \label{ss:SF}
We calculate the total water inventory of a planet is distributed between the planet's interior and surface (see \S \ref{s:SF}). The mineralogy of the interior affects the water storage capacity and likely impacts the transition zone water inventory. Here, we incorporate surface flooding to constrain the maximum oceans that rocky planets may have given their composition. 

Figure \ref{fig:SurfFlood} shows surface flooding as a function of Mg/Si and FeO content for a 1 M$_\oplus$ planet with a T$_m$ of 1600 K. Generally, as the oxygen availability within the mantle increases, the FeO content increases, which causes the surface water to increase. This is largely due to the method of water delivery via FeO + H$_2$. Planets with high Mg/Si and low FeO content (higher water storage capacities with lower water production) maintain the majority of their water inventory within their mantles.  At Mg/Si $\sim$1--1.3 and FeO  $\sim$10--15 wt\%, there is also a reduced amount of surface water that corresponds with the transition of olivine at lower pressures. Within the compositional parameter space considered in this study, this region has the highest water storage capacity. Above FeO $\sim$5 wt\%, most compositions have at least 1 Earth ocean of water, which increases rapidly with increasing FeO. We find a maximum of 8.5 surface oceans for a 1 M$_\oplus$ planet.

\subsubsection{Planet Topography}
We use hypsometric curves calculated from Solar System rocky planets to determine the influence of topography on surface flooding (see \S \ref{ss:mHypso})
The topography of the planet impacts the water coverage fraction of the surface and the seafloor pressures of the ocean basin. For all topographies, we find that all compositions have water coverage fraction of $>$ 0.5 above FeO $\sim$9 wt\%. Compositions that produce water coverage fractions similar to Earth (50--75\%), exist over a small region of FeO contents for a given Mg/Si ratio. Specifically, we find we find that changes of 1-3 wt\% of FeO significantly impact the water coverage fraction for all topographies.

For Earth-like topographies (Figure \ref{fig:SurfFlood}.A), we recover Earth with a FeO $\sim$8 wt\% and Mg/Si $\sim$1.2 with a water coverage fraction of $\sim$70\%. We find that $\sim$4.1 OM is the minimum water mass to submerge the surface for a 1 M$_\oplus$ planet with Earth-like topography. For the majority of compositions, we find that the surface is completely flooded at FeO $\sim$ 7--11 wt\%.  However, we also find three “islands" that do not exhibit complete surface flooding until FeO$\sim$13--15 wt \%.  These “islands" correspond to the highest water storage capacity compositions. Within these regions, the water coverage fraction varies from 90--99\%.

Venus-like and Mars topographies are shown in Figure \ref{fig:SurfFlood}.B and C, respectively. Given that their hypsometric curves have the largest range in elevation, they require significantly more water to submerge the surface, assuming an FeO content of Earth. For Venus-like topographies, 4.7 OM submerges the surface, whereas Mars-like topographies require $\sim$11.5 OM for complete submersion for a  1 M$_\oplus$ planet. Given that the maximum surface water that we find is $\sim$8.5 OM, a Mars-like topography would not be completely submerged at 1 M$_\oplus$. However, more massive planets do produce enough oceans to cover a Mars-like topography, which we will discuss further in \S \ref{s:KP}. However, Venus-like topographies do become completely covered above FeO= 13 wt\% at low Mg/Si (0.5--0.8) and above FeO= 20 wt\% at high Mg/Si.

A Mercury-like topography is more likely to occur for super-Earths than Earth-like rocky planets. Given that they likely have higher surface gravities, the resulting topography is flatter, similar to Mercury. Compared to Earth's topography the basin of the Mercury is shallow and the land becomes completely submerged with 2 OM of water (Figure \ref{fig:SurfFlood}.D). As a result, there is less variation in compositions necessary to submerge the surface completely. For all Mg/Si ratios, the land is completely covered at FeO $\sim$9-11 wt\%.

\subsection{Known Planets}\label{ss:KP}
To understand the water storage capacities and potential surface flooding of rocky planets, we extend our analysis to a sample of 689 known planets with host star abundances (see \S \ref{s:Method} and \S \ref{ss:CalculatingKP}). Here, we set their mantle potential temperatures to 1600K. We also assume an FeO = 8 wt\% similar to Earth unless otherwise noted due to the limited data on FeO content of known planet systems (see \S \ref{ss:FeO}).   

\subsubsection{Mantle Water Storage Capacity}\label{ss:KPWSC}
\begin{figure}[!t] 
\begin{center}
\includegraphics[width=\linewidth]{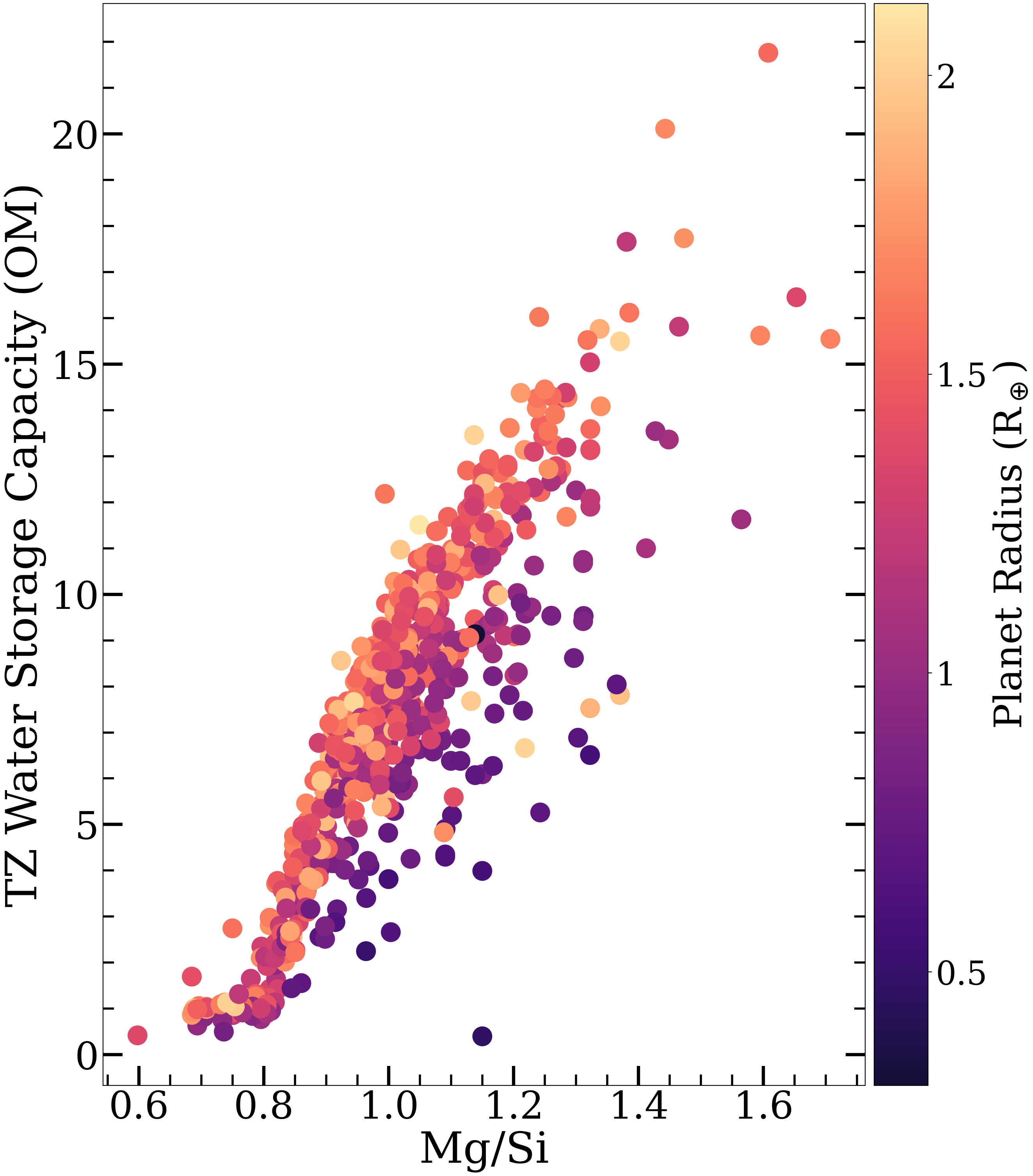}
\caption{Known Planets Water Storage Capacity as a function of Mg/Si. The colorbar corresponds to the planet radius. Here, we set the FeO to be 8 wt\% assuming the oxidation state of Earth.  }
\label{fig:KPvWaterStorage}
\end{center}
\end{figure}

Figure \ref{fig:KPvWaterStorage} shows the calculated water storage capacities of known rocky planets. For a given Mg/Si, the water storage capacity of the planet increases with planet radius. Within our planet sample, we find that the host star Mg/Si ranges from 0.59 -- 1.71 with the median being 1.01. With $\sim$55\% of planet host stars having a Mg/Si ratio above 1,  the majority of planets are within the high water storage capacity regime where high-pressure (Mg,Fe)$_2$SiO$_4$ forms. We find that the median water storage capacity of our sample is 7.8 OM, assuming a similar oxidation state (FeO content) to Earth. We find that with increasing Mg/Si, planet radius has a stronger affect with the differences in water storage capacity. For example, at Mg/Si= 0.8 the water storage capacity ranges from 0.5--1.5 OM, whereas at Mg/Si= 1 ranges from 2.5--8 OM.

\subsubsection{Surface Flooding \& Seafloor Pressures}\label{ss:KPSF}
Given that we do not have constraints on the FeO content of our planet sample, we explore two scenarios. We consider the water inventory using (1) an Earth-like FeO content of 8 wt\% and (2) a maximum water inventory assuming that each planet reaches the water solubility limit of 5.8 wt\% during its magma ocean phase (Miozzi et al 2025 (submitted)) .

Given an FeO content of 8 wt\%, we find the maximum number of surface oceans produced to be 35 OM, and the median to be 6.2 OM. The 95\% and 5\% quantiles are 22 OM and 0.05 OM, respectively. The large range of surface oceans is a strong function of planet radius and Mg/Si ratio. Planets below the 5\% quantiles have a higher median Mg/Si of $\sim$1.2 and a median radius of 0.7 $R_\oplus$, whereas above the 95\% quantile these planets have a lower median Mg/Si $\sim$0.9 and larger median radius of 1.7 $R_\oplus$. 

The topography impacts the distribution of the surface oceans, land mass, and whether complete surface flooding occurs. Assuming an FeO content of 8 wt\%, we find that Mars-like topography does not exhibit complete surface flooding for any planet in our sample. The median water coverage fraction is 98\%, which corresponds in the peak in the hypsometric curve (see Figure \ref{fig:Hyps}). While complete surface flooding does not occur, most planets only have 1-3\% land exposure. We find that with Venus and Earth-like topographies our planet sample begins to exhibit complete surface flooding, where 24\% and 35\% of planets have fully submerged surfaces, respectively. Above $\sim$1.5 $R_\oplus$ the majority of planets begin to have complete surface flooding for an Earth-like topography, whereas a Venus-like topography transitions to full submersion at $\sim$1.6 $R_\oplus$. For a Mercury-like topography, the 68\% of planets in our sample have complete surface flooding with the submersion radius of $\sim$1.2 $R_\oplus$ for most Mg/Si ratio. 

\begin{figure}[!t] 
\begin{center}
\includegraphics[width=\linewidth]{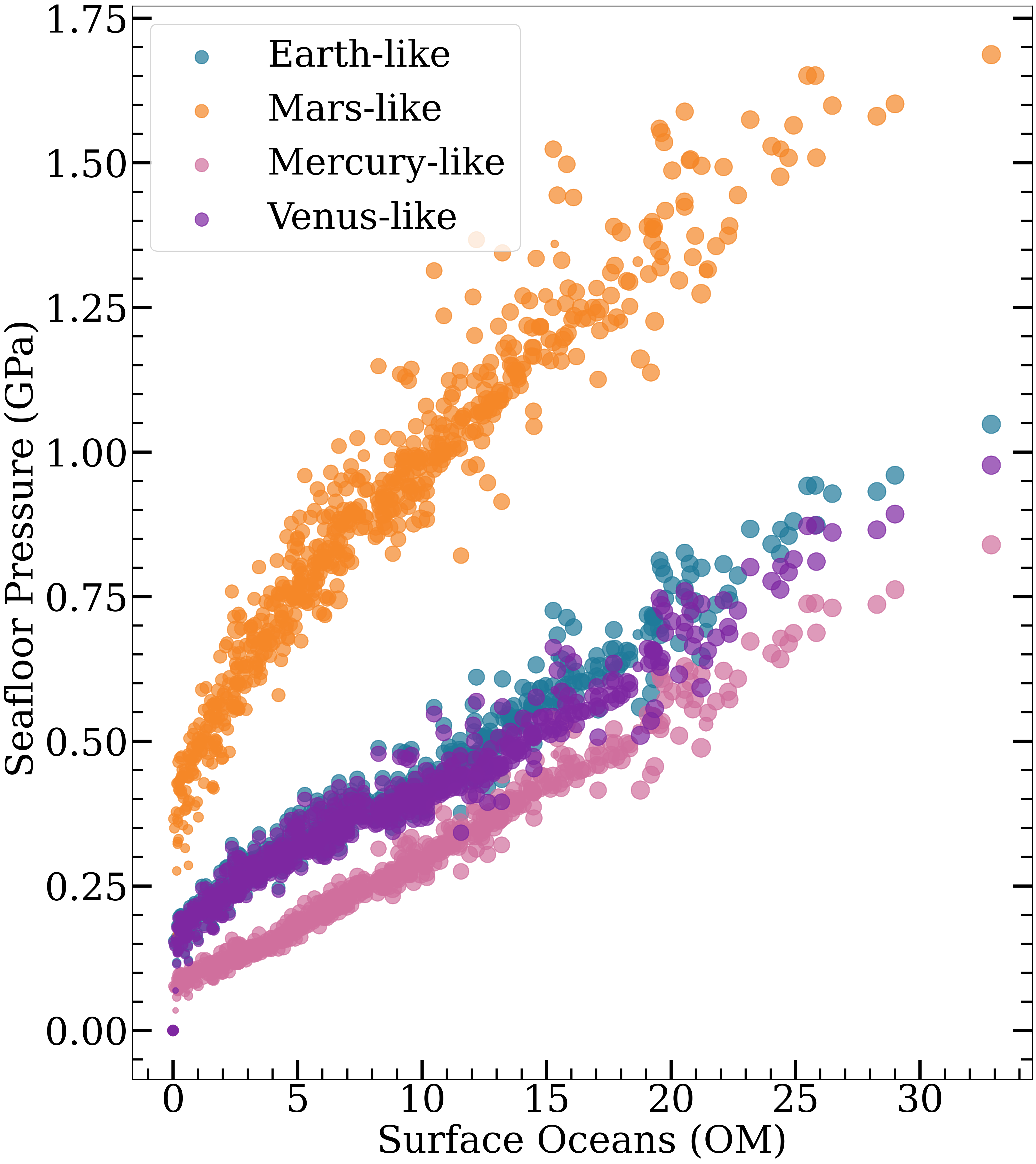}
\caption{Seafloor Pressures as a function of surface oceans. We show seafloor pressures for our known planet sample assuming an FeO content of 8 wt\% for differing topographies. We show Earth-like (teal), Mars-like (orange), Venus-like (purple), Mercury-like (pink) topographies. The size of the point corresponds to planet radius. }
\label{fig:seafloor}
\end{center}
\end{figure}

The corresponding seafloor pressures for our planet sample, assuming water inventories of an FeO=8 wt\% are shown in Figure \ref{fig:seafloor}. We show the entire planet sample for four solar system topographies. Within this regime, we find that the topography significantly impacts the seafloor pressures for a given OM. A Mercury-like topography has the smallest change in elevation, similar to the expected topography of super-Earths. The small change in elevation leads to the surface water mass being more evenly distributed. For this reason, we find that it has the lowest seafloor pressures. Given that high-pressure ice is thought to affect seafloor weathering and climate \cite[e.g.,][]{abbot2012,Kite2018,Chambers2020}, we note that Mercury-like topographies does not reach seafloor pressures above $\sim$1 GPa where high pressure ice forms at an equilibrium temperature of $\sim$300 K \cite[e.g.,][]{Journaux2013}. Similarly, we find that the majority of our planet sample, assuming Earth or Venus topographies have seafloor pressures below 1 GPa. In contrast, a Mars-like topography has the largest change in elevation and thus the highest seafloor pressures. However, only 24\% of our planet sample have seafloor pressures above 1 GPa assuming a Mars-Like topography. 

We explore the second scenario where a maximum water inventory is calculated assuming that each planet reaches the water solubility limit of 5.8 wt\% within the melt during its magma ocean phase. Here, we find the surface water to range from 14--2800 OM. However, the 5\% and 95\% quantiles are 82 OM and 1600 OM, respectively. Within this high OM regime, the Mg/Si has a negligible effect on reducing the surface water as the maximum water storage capacity of the mantle is $\sim$22 OM. Similarly, the seafloor pressure show negligible differences based on topography as they converge at at high OM. The seafloor pressures range from 0.59--59 GPa with the median being 12 GPa. At these pressures, high-pressure ice is likely to form even when accounting for salinity of the ocean \cite[e.g.,][]{Journaux2013}.





\section{Discussion}\label{s:Discuss}

\subsection{Water Storage Capacity of the Upper and Lower Mantle}\label{ss:WSCupperlower}
We consider the effect of composition on the water storage capacity of transition zone minerals, as it dominates the water storage capacity of the mantle of rocky planets \citep[e.g.,][]{Guimond2023}. The inclusion of water storage capacities of the upper and lower mantle will scale the water storage capacities higher. Compared to the transition zone the upper and lower mantle tend to be drier due to their mineralogy. The lower mantle composition is dominated by bridgmanite, where the water storage capacities are 2--3 orders of magnitude lower than wadsleyite and ringwoodite \citep{Panero2020}. For this reason, previous studies constraining the water storage capacities of the lower mantle find H$_2$O $\sim 0.15$ wt\%, corresponding to $\sim$1-3 OM \cite[e.g.][]{Litasov2003,Kaminsky2018}. However, some studies find that the lower mantle can have higher water storage capacities above $\sim 0.15$ wt\% \cite[e.g.][]{Fu2019}. Within the upper mantle, \cite{Guimond2023} found that on average 1 OM of water can be stored for a 1 M$_\oplus$ planet, but varies between $\sim$ 0.5-3 OM depending on the composition of the planet.

\subsection{FeO Constraints}\label{ss:FeO}
In this study, we show that variations in the FeO content similar to that of solar system have a significant impact on the water storage capacity and water production. However, the amount of FeO within the mantle is a reflection of the oxidation state of the mantle, and it is often considered using the oxygen fugacity \cite[e.g,][]{Guimond2023b}. The equilibrium reaction between Fe and FeO defines the reference fugacity commonly used when considering oxidation, which is express as

\begin{equation}
    Fe+ \frac{1}{2} O_2 \rightleftharpoons FeO
\end{equation}

Rocky planets within the solar system vary in their FeO content. Earth, Mars, Venus, and Mercury have an FeO content of $\sim$8 wt\% \cite[e.g.,][]{Workman2005}, $\sim$13 wt\% \citep{Khan2022}, $\sim$ 8--10 wt\% \citep{Surkov1984,Surkov1986}, and 1--4 wt\% \citep{Nittler2018}, respectively. However, including smaller solar system bodies, FeO content shows a strong correlation with mass. Specifically, increasing mass correlates to a decrease in the FeO content with Mercury being an outlier \citep{Johansen2023}. \cite{Johansen2023} suggests that the driving mechanism for the variation in the FeO content may be due to prolonged interaction with primordial ice caused by longer formation times. Ices can increase the FeO content within the planet. For this reason, we might expect super-Earths to be less oxidized than their Earth-like counterparts. In the context of this study, super-Earths having a lower FeO content would result in higher water storage capacities compared to Earth and less surface water, assuming that water production is dominated by the conversion of FeO+H$_2$. In addition to planet mass, the oxidation state is thought to be affected by the orbital distance from the sun due to the iron distribution, which is driven by the magnetic field in the protoplanetary disk \cite[e.g.,][]{McDonough2021}. Previous studies show a clear correlation with orbital distance, where planets are more oxidizing (higher FeO content) with increasing distance \citep{Carter2019,Charlier2019}. If FeO content is largely dominated by orbital distance, our results would suggest that rocky planets further from their host star would have higher water storage capacities in their mantle and a higher flooding potential than those on shorter orbits. 

The oxidation state of different planetary systems may vary compared to the solar system. Constraints on the planet oxidation state outside of the solar system is an emerging area of interest. \cite{Guimond2023b} demonstrates that the oxidation state (FeO content) for different planetary systems is likely to vary significantly compared to Earth. They show that increasing Mg/Si corresponds to up to a two order of magnitude increase in the oxygen fugacity of the system. This indicates that planets with high Mg/Si ratios may also have a higher FeO content. As a result, both the water storage capacity and initial water would significantly increase compared to the result above. Assuming that the majority of the water inventory is accreted via interaction with the H$_2$ atmosphere, high Mg/Si ratios would likely also have high water mass fractions and complete surface flooding. 

In addition to theoretical studies, the oxidation state of planets may be constrained through the study of polluted white dwarfs. In \cite{Doyle2019}, the authors consider a sample of 6 polluted white dwarfs to understand the oxygen fugacity of rocky material that has accreted into their atmospheres, which are generally composed of hydrogen. They found that the oxidation state is similar to that of chondrites within the solar system. However, the abundances within the atmospheres of polluted white dwarfs vary as material accretes over time due to their short diffusion time scales \citep{Brouwers2023}. Compared to main sequence stars, polluted white dwarfs follow significantly different distributions of Fe/Ca, indicating the ejection of mantle material \citep{Brouwers2023}. The Fe/Ca distribution of polluted white dwarfs makes it unclear whether the oxidation state of the planetary material can accurately be determined with observations of a polluted white dwarf alone. Stars also vary chemically over Galactic space and time, which adds further complexity to inferring the planetary material oxidation state. Without understanding its initial chemical composition, it is difficult to constrain the mantle material using polluted white dwarfs. However, white dwarfs with a main-sequence wide binary companion can be used to constrain the initial compositions, because binary systems form out of the same interstellar gas and have similar ages and initial compositions. Given that several studies \cite[e.g,][]{Aguilera2025} have constrained refractory elements within polluted white dwarfs, it is likely that the FeO content can be better constrained for planets using white dwarfs in future studies.


\subsection{Mantle Water Cycling}\label{ss:WaterCycling}

To understand the long-term evolution of surface conditions and their water inventory, further studies are necessary to constrain water cycling on rocky planets. On Earth, the estimate for the amount of surface water lost to the mantle is 20\% per billion years with a cooling rate of $\sim$100 K/Ga \citep{Andrault2022}. However, planets with different compositions to Earth will likely have varying surface water loss due to the role of planet composition on the water storage capacity of the mantle.  We show that different Mg/Si ratios are more sensitive to changes in temperature due the planet's mineralogy similar to previous studies \cite[e.g.,][]{Panero2020, Guimond2023}.  The amount of FeO will likely change the rate at which a planet with surface water is lost to the mantle. At high FeO contents ($\gtrsim$11 wt\%),  olivine transitions to wadsleyite at lower pressures, which significantly increases the water storage capacity. Planets with high Mg/Si ($\gtrsim$1.2) will also likely sequester surface water at a faster rate than Earth over their evolution if they exhibit similar dynamics due to the drastic increase in the water storage capacities of (Mg,Fe)$_2$SiO$_4$. At high Mg/Si ($\gtrsim$1.2) compositions are more sensitive to changes in temperature and have an increase in the water storage capacity of 3 times the OM of low Mg/Si compositions. It is clear that further studies are necessary to constrain how the water inventory of a rocky planets may vary between the surface and mantle over time, accounting for temperature, composition, and geodynamics. Understanding the water cycle of planets will help constrain if and how long a planet may sustain a stable climate.

\subsection{Role of Water and Topography on Climate Stability}\label{ss:StableClimates}




A stable climate and surface water are often thought to be essential for habitability \cite[e.g.,][]{Spiegel2008,Foley2016, Kite2018}. The long-term climate stability of a planet is largely a function of the atmospheric CO$_2$ content and its regulation, which is impacted by stellar instellation, surface water, and tectonic regime (i.e, plate tectonics, stagnant lid, etc.) \citep{abbot2011,abbot2012,FOLEY2012, Kodama2015,Foley2016,Kodama2019,Chambers2020, Hayworth2020}. Silicate weathering buffers atmospheric CO$_2$ levels and thus acts to stabilize the climate \citep{Willenbring2010}. On Earth, the carbonate-silicate (C-S) cycle draws down CO$_2$ from the atmosphere through silicate weathering and sequesters CO$_2$ in carbonate sediments that subduct at plate boundaries into the mantle \citep{Walker1981, Kasting1993, Berner2004}. The net drawdown of CO$_2$ depends on the weathering rate of Ca-Mg silicate rocks and the resulting supply of calcium that combines with dissolved CO$_2$ in surface oceans to form carbonate sediments\cite[e.g.,][]{Berner2004}. The negative feedback between global temperatures and silicate weathering rates regulates atmospheric CO$_2$, and thus, the climate. Higher CO$_2$ produces warmer climates and is associated with higher precipitation and increased silicate weathering that draws down CO$_2$ \citep{Walker1981,Berner1983, Berner2004}. In colder climates, outgassing outpaces the lowered weathering sink, allowing CO$_2$ to rise and maintain climate stability \citep{Walker1981, Berner2004, Maher2014}.  

Silicate weathering may not be enough to regulate atmospheric CO$_2$, especially at more extreme instellations. At sufficiently low stellar flux, planets with surface oceans enter into a snowball state where oceans are entirely ice-covered \citep{abe2011,Kadoya2014, Menou2015}. Conversely, planets with surface oceans can also enter into a runaway greenhouse state. Evaporation driven by high stellar flux can leads to complete surface water loss, leading to an inhospitable  environment \citep{Kasting1988, goldblatt2013}. 
The evaporation of liquid water creates water vapor
within the atmosphere, resulting in steam-dominated atmosphere driving a positive feedback  that warms the planet. The increase in water vapor traps thermal radiation in the atmosphere, causing temperatures to continue to increase evaporation \cite[e.g.,][]{goldblatt2013,Gomez-Leal2018}. However, the water inventory and the distribution of the land influence the stellar flux threshold that triggers a runaway greenhouse and snowball state \citep{Kodama2018, Kodama2019, Kodama2021}.

On a planet with partial flooding, the threshold of stellar flux that triggers a runaway greenhouse or snowball state is controlled by the distribution of land and water (i.e, planet topography) \citep{Kodama2018, Kodama2019,Way2021,  Kodama2021, Zhao2021,Glaser2025}.  Water coverage fraction and land distribution modulate planetary climate by changing planetary surface albedo, surface heat capacity, and the sources of atmospheric moisture \cite[e.g.][]{Kodama2018,Way2021}. The instellation threshold for a runaway greenhouse state extends to higher instellation if the land distribution enables atmospheric moisture transport to maintain a dry tropical region \citep{Kodama2018}. Similarly, the instellation threshold for a snowball state is dependent on the land distribution.  Given the same water coverage fraction, changes in the land distribution may result in a cooler climate and thus higher instellation in which the planet enters a snowball state \cite[e.g.,][]{Kodama2021}. Therefore, topography and water coverage fraction are strong controls on the climate state of a planet. 

Topographies similar to those on Earth, Venus, and Mars are more likely to result in partial flooding due to their large ocean basins (see \S \ref{s:SF}). On partially flooded planets, climate stability depends on the efficiency of silicate weathering. Erosion, precipitation, runoff, and the locations of atmospheric moisture sources relative to those of weatherable land surfaces largely control weathering rates, influencing climate regulation \citep{Berner2004, West2012, Maher2010, Maher2014,Baum2022}.

For flatter, Mercury-like topographies, our results suggest that the majority of rocky planets will be completely flooded, assuming an Earth-like FeO content or a maximum H$_2$O solubility limit of 5 wt\%.  Depending on the amount of surface water, ocean planets (i.e., planets with complete surface flooding) may have a narrower range of habitable surface conditions \citep{abbot2012, Kite2018, Nakayama2019, Chambers2020,Hayworth2020}. Ocean planets with seafloor pressures $\lesssim$1 GPa may regulate their climate efficiently via seafloor weathering as an alternative mechanism to buffer against changes in climate due to stellar flux and outgassing \citep{krissansen-totton2017,Chambers2020,Hayworth2020}. Seafloor weathering is a process by which basaltic rock releases Ca$^{2+}$ that reacts with dissolved CO$_2$ to precipitate carbonate in pore spaces \citep{krissansen-totton2017, Coogan2013,Coogan2015}. The long-term CO$_2$ sink of deposition and subduction of carbonate sediments depends on a combination of seafloor spreading rates, crustal pore space, and the timescale of atmosphere-deep ocean CO$_2$ equilibration \cite[e.g.,][]{Chambers2020}. Ocean planets with seafloor weathering may be more effective at buffering against changes in stellar flux via seafloor weathering compared to partially flooded planets \citep{Hayworth2020}. However, they are more sensitive to changes in temperature due to volcanic outgassing compared to partially flooded planets where continental weathering dominates \citep{Hayworth2020}.   Ocean planets with seafloor pressures $\gtrsim$1 GPa likely form high-pressure ices, which impact seafloor weathering \citep{Abbot2016,Kite2018, Nakayama2019}. If the high-pressure ice layer is solid, the ocean planet may maintain a stable climate for up to $\sim$2 Gyr, depending on the instellation \citep{Kite2018}. These planets will likely end up with an accumulation of atmospheric CO$_2$, leading to a runaway greenhouse state \citep{Alibert2014,Kitzmann2015}. However, ocean planets with a liquid-solid mixture of water and ice near mid-ocean ridges the planet will likely enter into a snowball state regardless of instellation \citep{Nakayama2019}. Ocean planets reaching the snowball state may enter a snowball or limit cycle where  the climate transitions from a snowball to warm climate. In this scenario, CO$_2$ from interior degassing accumulates within the atmosphere above a CO$_2$ threshold to enter into a snowball state, causing the planet to warm \citep{Tajika2008,Kadoya2014, Abbot2016,Haqq2016,Kadoya2019}.

Drier to fully-dry planets ($\lesssim$0.5 OM) despite having exposed land for active continental weathering, may also lack a mechanism for long-term (solid) carbon storage and recycling if atmospheric moisture for precipitation is insufficient or carbonates cannot be deposited, thermally decomposed, and returned to the atmosphere as CO$_2$ \citep{abe2011, Glaser2025}. Drier planets are less likely to experience catastrophic climate instabilities such as a planetary snowball or runaway greenhouse state but may still require a long-term climate regulation mechanism to remain habitable if o remain habitable if outgassed CO$_2$ accumulates in the atmosphere over long timescales \citep{abe2011, Kodama2019}.

\section{Summary \& Conclusions} \label{s:summary and conclusions}

In this paper, we constructed a homogeneous rocky planet sample with host star abundances to constrain the water inventory as a function of Mg/Si and FeO content. We provide the first exploration of the influence of planet composition on the maximum surface flooding potential of rocky planets and the impact of topography on seafloor pressures. We quantify the impact of FeO content on mantle transition zone water storage capacities. We also provide a homogeneous catalog of 689 rocky planets with host star abundances from APOGEE or GALAH. 

With the Habitable Worlds Observatory's main objective to identify and directly image 25 potentially habitable worlds.  Given that planets with habitable conditions likely require surface water, this study determines compositional constraints for surface water.

From this study, we present our primary conclusions:

\begin{enumerate}
    \item We find that FeO content in the mantle significantly impacts the amount of water that may be stored within the mantle. Higher FeO content (above $\sim$11 wt\%) increases the volume of the transition zone, which has implications for the long-term evolution of the planet (\S \ref{s:WSC}).

    \item Flatter topographies, as expected for super-Earths, have a smaller range of compositions conducive to exhibit silicate weathering (Figure \ref{fig:SurfFlood}). Therefore, these planets may be reliant on seafloor weathering to stabilize their climate or may be habitable for less time depending on their water mass (\S \ref{ss:StableClimates}).
    
    \item As a planet cools, it will increase it's water storage capacity. Mg/Si ratios above 1.2 show an increase of 5 times the water storage capacity in the transition zone at 1500K compared to 2000K. Given that Earth loses $\sim$20 \% of its water per billion years, planets that vary in composition compared to Earth will likely have different rates of water loss to the mantle. Further studies that include thermal and dynamical evolution in addition to planet composition are necessary to determine how long a planet might maintain its surface water  (\S \ref{ss:Temp}) . 

    \item We provide mantle water storage capacities assuming an Earth-like FeO content (8 wt\%) and using their host star abundances to constrain planet composition for 689 super-Earths (\S \ref{ss:KP}).

    \item We provide scaling function to calculate water storage capacities of known planets using their radius and Mg/Si and FeO content (Equation \ref{eq:WSC}, \S \ref{ss:KPWSC}).
    
\end{enumerate}

\begin{acknowledgments}
We thank the anonymous reviewer for their comments that have improved this paper. KMB acknowledges support from NASA through the NASA Hubble Fellowship grant
HST- HF2- 51497 awarded by the Space Telescope Science Institute, which is operated by the Association of Universities for Research in Astronomy, Inc., for NASA, under contract NAS5-26555. We thank Nicole Wahlstrom for her efforts on land depression caused by surface flooding.
\end{acknowledgments}

 \software{ \texttt{ExoPlex}~\citep{2018NatAs...2..297U,Unterborn2019, Unterborn2023}}

\clearpage

\begin{appendix}
\setcounter{table}{0}
\renewcommand{\thetable}{A\arabic{table}}

\section{Planet Sample}
We show the first 50 columns of super-Earth planet parameters and water storage capacities in Table~\ref{tab:sE}.
\label{Appendix}
\startlongtable
\movetableright=-1in
\begin{deluxetable*}{|c|c|c|c|c|c|c|c|c|c|}
    \tabletypesize{\scriptsize}
    \tablecaption{Super-Earth Parameters\label{tab:sE}}
    \tablecolumns{9}
    \tablehead{
   \colhead{Planet} & \colhead{P (days)} & \colhead{Radius(R$_\oplus$)} & \colhead{ Mass(M$_\oplus$)}&\colhead{M-R Source } &\colhead{Fe/Mg } &\colhead{ Mg/Si } &\colhead{Ca/Mg$^a$ } &\colhead{Al/Mg$^b$ }&\colhead{W$_{TZ}$ (OM)$^c$}
}
    \startdata
    \hline
        CoRoT-7 b&0.85 &1.68$\pm$0.11&6.06$\pm$0.65&1& 0.57$\pm$0.03& 1.04$\pm$0.11& 0.05$\pm$0.03&0.09$\pm$0.04 & 7.54\\
      EPIC 220674823 b&0.57 &1.68 $\pm$0.04&7.80$^{0.71}_{-0.70}$& 2 & 0.61$\pm$0.03&1.02$\pm$0.09&0.05$\pm$0.04&0.12$\pm$0.05&7.54\\
      HD 3167 b& 0.96&1.67$^{0.17}_{-0.10}$& 4.97$^{0.24}_{-0.23}$& 3&0.30$\pm$0.06& 1.24$\pm$0.26& 0.04$\pm$0.03& 0.11$\pm$0.10 &16.02	 \\
      K2-111 b&5.35 &1.82$^{0.11}_{-0.09}$&5.58 $^{0.74}_{-0.73}$& 3&0.15$\pm$0.04& 1.19$\pm$0.64&0.04$\pm$0.06&0.10$\pm$0.06 & 12.35\\
      K2-138 b& 2.35&1.51$^{0.11}_{-0.08}$&3.10$\pm$1.05& 4&"0.61$\pm$0.03& 1.04$\pm$0.08& 0.06$\pm$0.05&0.10$\pm$0.06 &7.20\\
      K2-141 b& 0.28&1.51$\pm$0.05&4.97$^{0.35}_{-0.34}$&3&"0.68$\pm$0.04& 1.12$\pm$0.08& 0.07$\pm$0.06& 0.09$\pm$0.09&9.04\\
      K2-216 b& 2.17&1.75$^{0.17}_{-0.10}$&8.00$\pm$1.60& 5&0.44$\pm$0.03&1.28$\pm$0.04&0.05$\pm$0.04&0.08$\pm$0.05& 11.68\\
      K2-265 b& 2.37&1.71$^{0.11}_{-0.08}$&7.34$^{1.43}_{-1.40}$& 6&"0.55$\pm$0.03& 1.07$\pm$0.09& 0.07$\pm$0.04&0.12$\pm$0.04& 8.93\\
        K2-32 e& 4.35& 1.21$\pm$0.05& 2.10$^{1.30}_{-1.10}$&7& 0.46$\pm$0.03&1.19$\pm$0.09&0.05$\pm$0.05& 0.09$\pm$0.06&9.10\\
      K2-38 b&4.02 &1.66$\pm$0.10 &7.70$^{1.20}_{-1.10}$ & 3& 0.61$\pm$0.02 &0.90$\pm$0.22&0.05$\pm$0.03&0.14$\pm$0.04& 5.08\\
        Kepler-10 b& 0.84&1.47$^{0.03}_{-0.02}$&3.26$\pm$0.30 &8& 0.35$\pm$0.04&1.11$\pm$0.28&0.06$\pm$0.05&0.12$\pm$0.06& 10.45\\
      Kepler-100 b& 6.89&1.32$\pm$0.04&7.34$\pm$3.20&9& "0.60$\pm$0.03&0.87$\pm$0.26&0.05$\pm$0.03&0.13$\pm$0.04& 3.10\\
      Kepler-102 d& 10.31&1.15$\pm$0.06&3.00$\pm$1.30& 3&0.50$\pm$0.04& 1.15$\pm$0.08 &0.05$\pm$0.05&0.08$\pm$0.08& 9.24\\
        Kepler-11 b& 10.3& 1.80$^{0.03}_{-0.05}$&1.90$^{1.40}_{-1.00}$&10&0.65$\pm$0.03& 0.95$\pm$0.14&0.06$\pm$0.04&0.11$\pm$0.05&5.04\\
      Kepler-114 b& 5.19&1.02$\pm$0.11&6.80$^{4.30}_{-3.60}$& 11 & 0.62$\pm$0.04& 1.04$\pm$0.08&0.06$\pm$0.05&0.10$\pm$0.08&7.12\\
      Kepler-114 c& 8.04&1.60$\pm$0.18&2.80$\pm$0.60&12&0.62$\pm$0.04&1.04$\pm$0.08&0.06$\pm$0.05&0.10$\pm$0.08&7.93\\
      Kepler-128 b& 15.00&1.42$\pm$0.04&3.79$^{0.76}_{-0.66}$&13& 0.71$\pm$0.03& 0.81$\pm$0.25&0.06$\pm$0.04&0.12$\pm$0.04& 3.33\\
      Kepler-128 c& 22.80&1.52$\pm$0.05&3.38$^{0.67}_{-0.59}$&13& 0.71$\pm$0.03& 0.81$\pm$0.25&0.06$\pm$0.04&0.12$\pm$0.04&3.35\\
      Kepler-131 c& 25.52&0.84$\pm$0.07&8.25$\pm$5.90& 9& 0.72$\pm$0.02&0.94$\pm$0.10&0.06$\pm$0.03&0.11$\pm$0.04&5.75\\
      Kepler-138 b& 10.31&0.64$\pm$0.02&0.07$\pm$0.02&14& 0.87$\pm$0.03& 1.00$\pm$0.08&0.07$\pm$0.05& &2.66\\
      Kepler-138 c& 13.78&1.51$\pm$0.04&2.30$^{0.60}_{-0.50}$& 14&0.87$\pm$0.03& 1.00$\pm$0.08&0.07$\pm$0.05& &6.04\\
      Kepler-138 d& 23.09&1.51$\pm$0.04&2.10$^{0.60}_{-0.70}$& 14& 0.87$\pm$0.03&1.00$\pm$0.08&0.07$\pm$0.05& &6.00\\
        Kepler-18 b& 3.5& 2.00$\pm$0.10&6.90$\pm$3.40& 15& 0.62$\pm$0.04&0.93$\pm$0.17&0.05$\pm$0.05&0.11$\pm$0.07& 5.78\\
      Kepler-197 c& 10.35&1.23$\pm$0.04&5.30$^{3.30}_{-2.90}$& 16&0.41$\pm$0.03& 1.02$\pm$0.29& 0.05$\pm$0.05& 0.09$\pm$0.05&7.63\\
      Kepler-1972 b&7.54& 0.80$\pm$0.04& 2.02$^{0.56}_{-0.62}$&17&0.66$\pm$0.03& 0.90$\pm$0.18& 0.06$\pm$0.04& 0.11$\pm$0.05&2.52\\
      Kepler-1972 c&11.33& 0.87$\pm$0.05& 2.11$^{0.59}_{-0.65}$&17&0.66$\pm$0.03& 0.90$\pm$0.18& 0.06$\pm$0.04& 0.11$\pm$0.05&2.79\\
      Kepler-21 b&2.79& 1.64$\pm$0.02& 7.50$\pm$1.30&3&0.93$\pm$0.03& 0.75$\pm$0.19& 0.07$\pm$0.04& 0.13$\pm$0.04&2.74\\
      Kepler-23 b&7.1& 1.64$\pm$0.05& 2.56$^{0.43}_{-0.40}$&13&0.58$\pm$0.04& 0.96$\pm$0.18& 0.05$\pm$0.05& 0.12$\pm$0.06&5.71\\
      Kepler-323 c&3.55& 1.57$\pm$0.04& 6.80$^{3.40}_{-3.20}$&3&0.55$\pm$0.03& 1.00$\pm$0.16& 0.05$\pm$0.04& 0.09$\pm$0.04&7.11\\
      Kepler-338 e&9.34& 1.56$\pm$0.07& 8.50$^{7.20}_{-6.30}$&11&0.66$\pm$0.04& 0.90$\pm$0.18& 0.06$\pm$0.07& 0.10$\pm$0.08&4.48\\
      Kepler-345 b&7.42& 0.80$\pm$0.10& 0.50$\pm$0.30&16&0.94$\pm$0.03& 1.03$\pm$0.15& 0.08$\pm$0.04& 0.08$\pm$0.05&4.25\\
      Kepler-345 b&9.39& 1.30$\pm$0.10& 2.20$\pm$0.90&16&0.94$\pm$0.03& 1.03$\pm$0.15& 0.08$\pm$0.04& 0.08$\pm$0.05&6.69\\
      Kepler-36 b&13.87& 1.50$^{0.06}_{-0.05}$& 3.83$^{0.11}_{-0.10}$&18&0.59$\pm$0.05& 0.93$\pm$0.20& 0.06$\pm$0.07& 0.10$\pm$0.08&5.50\\
      Kepler-48 b&4.78& 1.88$\pm$0.10& 3.94$\pm$2.10&9&0.44$\pm$0.02& 1.11$\pm$0.17& 0.05$\pm$0.03& 0.10$\pm$0.04&9.01\\
      Kepler-595 c&12.39& 1.01$\pm$0.02& 3.30$^{1.70}_{-1.00}$&19&0.51$\pm$0.03& 1.12$\pm$0.10& 0.05$\pm$0.04& 0.10$\pm$0.06&8.96\\
      Kepler-65 b&2.15& 1.44$^{0.04}_{-0.03}$& 2.40$^{2.40}_{-1.60}$&20&1.10$\pm$0.02& 0.68$\pm$0.20& 0.07$\pm$0.03& 0.13$\pm$0.04&1.70\\
      Kepler-65 d&8.13& 1.59$^{0.04}_{-0.04}$& 4.14$^{0.79}_{-0.80}$&20&1.10$\pm$0.02& 0.68$\pm$0.20& 0.07$\pm$0.03& 0.13$\pm$0.04&1.72\\
      Kepler-93 b&4.73& 1.48$\pm$0.02& 4.66$\pm$0.53&3&0.52$\pm$0.03& 1.01$\pm$0.18& 0.05$\pm$0.04& 0.10$\pm$0.04&7.28\\
      Kepler-98 b&1.54& 1.99$\pm$0.22& 3.55$\pm$1.60&9&0.63$\pm$0.02& 0.96$\pm$0.13& 0.06$\pm$0.03& 0.09$\pm$0.04&5.91\\
      Kepler-99 b&4.6& 1.48$\pm$0.08& 6.15$\pm$1.30&9&0.57$\pm$0.02& 1.06$\pm$0.08& 0.06$\pm$0.03& 0.09$\pm$0.04&8.49\\
      LHS 1815 b&3.81& 1.09$\pm$0.06& 1.58$^{0.64}_{-0.60}$&21&0.39$\pm$0.03& 1.41$\pm$0.03&& &11.00\\
      TOI-1685 b&0.67& 1.47$\pm$0.05& 3.03$^{0.33}_{-0.32}$&22&0.87$\pm$0.03& 0.95$\pm$0.03&& &5.30\\
      TOI-1798.02&0.44& 1.40$^{0.07}_{-0.06}$& 5.60$^{0.80}_{-0.70}$&23&0.51$\pm$0.03& 1.10$\pm$0.10& 0.05$\pm$0.04& 0.09$\pm$0.04&5.59\\
      TOI-286 b&4.51& 1.42$\pm$0.10& 4.53$\pm$0.78&24&0.52$\pm$0.03& 1.01$\pm$0.18& 0.05$\pm$0.04& 0.10$\pm$0.04&7.02\\
      TOI-561 b&0.45& 1.40$\pm$0.03& 2.02$\pm$0.23&25&0.20$\pm$0.03& 1.27$\pm$0.44& 0.04$\pm$0.05& 0.09$\pm$0.05&12.78\\
      WASP-47 e&0.79& 1.83$\pm$0.02& 9.00$^{0.60}_{-0.40}$&26&0.57$\pm$0.02& 0.87$\pm$0.28& 0.05$\pm$0.03& 0.11$\pm$0.04&3.86\\
      EPIC 249893012 b&3.6& 1.95$^{0.09}_{-0.08}$& 8.75$^{1.09}_{-1.08}$&27&0.55$\pm$0.06& 1.18$\pm$0.06&0.1& 0.11$\pm$0.12&9.98\\
      K2-131 b&0.37& 1.69$^{0.09}_{-0.06}$& 7.90$\pm$1.30&3&0.65$\pm$0.04& 0.94$\pm$0.14&0.07& 0.06$\pm$0.11&5.76\\
      K2-199 b&3.23& 1.73$^{0.05}_{-0.04}$& 6.90$\pm$1.80&28&0.22$\pm$0.03& 1.26$\pm$0.38&0.06& 0.06$\pm$0.08&12.72\\
      K2-291 b&2.23& 1.59$^{0.10}_{-0.07}$& 6.49$\pm$1.16&29&0.63$\pm$0.03& 1.13$\pm$0.04&0.05& 0.09$\pm$0.06&9.06\\
      \ldots   & \ldots & \ldots & \ldots & \ldots& \ldots& \ldots& \ldots& \ldots& \ldots\\ 
    \enddata
        \tablecomments{ 
        We calculate host star molar ratios that were derived from chemical abundances within APOGEE and GALAH using solar normalization from \cite{Lodders2010}. The uncertainties quoted are conversions from APOGEE and GALAH to molar uncertainties using the methodology outlined in \cite{Hinkel2014}. 
        \tablenotetext{a}{Blanks values indicate that quality [Ca/Fe] was not available within our stellar sample.}
        \tablenotetext{b}{Blanks values indicate that quality [Al/Fe] was not available within our stellar sample.}
\tablenotetext{c}{We show the water storage capacities within the transition zone assuming a mantle potential temperature of 1600 K.}
}

        \tablerefs{
 (1) \citet{Barros2014}, 
 (2) \citet{Guenther2024}, 
 (3) \citet{Bonomo2023}, 
 (4) \citet{Lopez2019}, 
 (5) \citet{Persson2018}, 
 (6) \citet{Thygesen2023}, 
 (7) \citet{Lillo-Box2020},
 (8) \citet{Bonomo2025},
 (9) \citet{Marcy2014},
 (10) \citet{Lissauer2013}, 
 (11) \citet{Hadden2014}, 
 (12) \citet{Xie2014},
 (13) \citet{Leleu2023}, 
 (14) \citet{Piaulet2023}, 
 (15) \citet{Cochran2011},
 (16) \citet{Rowe2014}, 
 (17) \citet{Leleu2022},
 (18) \citet{Vissapragada2020},
 (19) \citet{Yoffe2021},
 (20) \citet{Mills2019}, 
 (21) \citet{Luque2022}, 
 (22)\citet{Burnt2024}, 
 (23) \citet{Crossfield2025},
 (24) \citet{Hobson2024},
 (25) \citet{Piotto2024}, 
 (26) \citet{Nascimbeni2023},
 (27) \citet{Hidalgo2020},
 (28) \citet{Akana2021},
 (29) \citet{Kosiarek2019}}
\end{deluxetable*}

\end{appendix}

\bibliographystyle{aasjournalv7}
\bibliography{Bib}

\end{document}